\documentclass[aps,showpacs,twocolumn]{revtex4-1}
\usepackage{epsfig}
\usepackage{graphicx,color,dcolumn}
\usepackage{epstopdf}
\usepackage{amsmath,amssymb}
\usepackage{multirow}
\usepackage{diagbox}
\usepackage{changes}
\usepackage{booktabs}
\usepackage{threeparttable}
\usepackage{subfigure}
\usepackage{hyperref}

\newcommand{\ty}{Yue Tan}

\newcommand{\hhx}{Hongxia Huang}
\newcommand{\jlp}{Jialun Ping}

\newcommand{\nj}{Nanjing}

\begin{document}
\title{ Equivalence among color singlet, color octet and diquark structure in a chiral quark model}
\author{\ty}
\email[E-mail: ]{tanyue@ycit.edu.cn}
\affiliation{Department of Physics, Yancheng Institute of Technology, Yancheng 224000, P. R. China}

\author{Xuejie Liu}
\email[]{1830592517@qq.com}
\affiliation{School of Physics, Henan Normal University, Xinxiang 453007, P. R. of China}

\author{Xiaoyun Chen}
\email[]{xychen@jit.edu.cn}
\affiliation{College of Science, Jinling Institute of Technology, Nanjing 211169, P. R.China}

\author{Yuheng Wu}
\email[E-mail: ]{191002007@njnu.edu.cn  }
\affiliation{Department of Physics, \nj~ Normal University, \nj~ 210023, P.R. China}

\author{\hhx}
\email[E-mail: ]{\hhx @njnu.edu.cn (Corresponding author) }
\affiliation{Department of Physics, \nj~ Normal University, \nj~ 210023, P.R. China}
\author{\jlp}
\email[E-mail: ]{jlping@njnu.edu.cn (Corresponding author)}
\affiliation{Department of Physics, \nj~ Normal University, \nj~ 210023, P.R. China}
\date{\today}

\begin{abstract}
Since the quark model was put forward, theoretical researchers have always attached great importance to the study of hidden color channels (including color octets and diquark structure). Because of the influence of color Van der waals forces, the hidden color channel itself has strong attraction, which provides a dynamic mechanism for the formation of resonance state or bound state. In this paper, taking the $T_{cc}$ system as an example, under the framework of multi-Gaussian expansion method, a set of relatively complete color singlets (that is, the ground state of the color singlet plus its corresponding higher-order component) is used to replace the contribution of the color octet. Similarly, we endeavor to replace the diquark structure with a relatively complete set of molecular states, encompassing both the ground state and excited states. Our results demonstrate that the color octet structure can be effectively replaced by a set of relatively complete color singlet bases, while the diquark structure cannot be entirely substituted by an equivalently comprehensive set of molecular state bases.
\end{abstract}

\maketitle

\section{Introduction} \label{introduction}
In the traditional quark model, mesons are composed of $q\bar{q}$ and baryons are composed of $qqq$. Under this framework, the traditional quark model successfully explained a large number of experimental data at that time and predicted the $\Omega$ particle. However, the basic theory of strong interaction, QCD, does not prohibit the existence of multiple quarks. As long as the number of quarks satisfies the formula, $q^{m}\bar{q}^n, m-n=3k,m+n>3$. In fact, since 2003, many possible multi-quark states, such as $X(3872)$ \cite{Choi:2003ue}, $P_c$ states \cite{LHCb:2015yax}, and so on, have been reported by various experimental groups.
Theoretically, there are varying perspectives: some contend that the predominant constituent of exotic states is the color octet structure \cite{Tan:2022pzi}, others propose that diquarks may play a leading role in exotic states \cite{Wang:2019veq}, and there are also those who argue that experimentally observed exotic states near the threshold are likely color singlet molecular states \cite{Ji:2022uie,Guo:2017jvc}.

Recently, the LHCb collaboration reported that the mass of $T_{cc}$ is about 1 MeV lower than the threshold $DD^{*}$ \cite{LHCb:2021auc}. However, within the framework of the chiral quark model, if only molecular state structures including color-singlet and color-octet states are taken into account, the calculated results align well with experimental findings. On the other hand, when diquark structures are considered, the lowest eigenenergy obtained is 100 MeV lower than the $DD^{*}$ threshold, significantly below the experimentally observed results \cite{Tan:2020ldi,Deng:2021gnb,Chen:2021tnn,Jin:2020yjn,Vijande:2003ki,Luo:2017eub,Yang:2019itm}. If $T_{cc}$ with energy of 3875 MeV is confirmed by other collaboration, which may pose a challenge to the chiral quark models. Hence, a profound understanding of the relationships among color singlet, color octet, and diquark structures stands as a crucial task within the current framework of the chiral quark model. The primary objective of this paper is to investigate the relationship between color singlet and color octet states, and diquarks, with the $T_{cc}$ system as a case study.

\begin{figure}[h]
\vspace{-0.1cm}
\setlength {\abovecaptionskip} {+0.2cm}
\setlength {\belowcaptionskip} {+0.1cm}
  \centering
  \includegraphics[width=8cm,height=3cm]{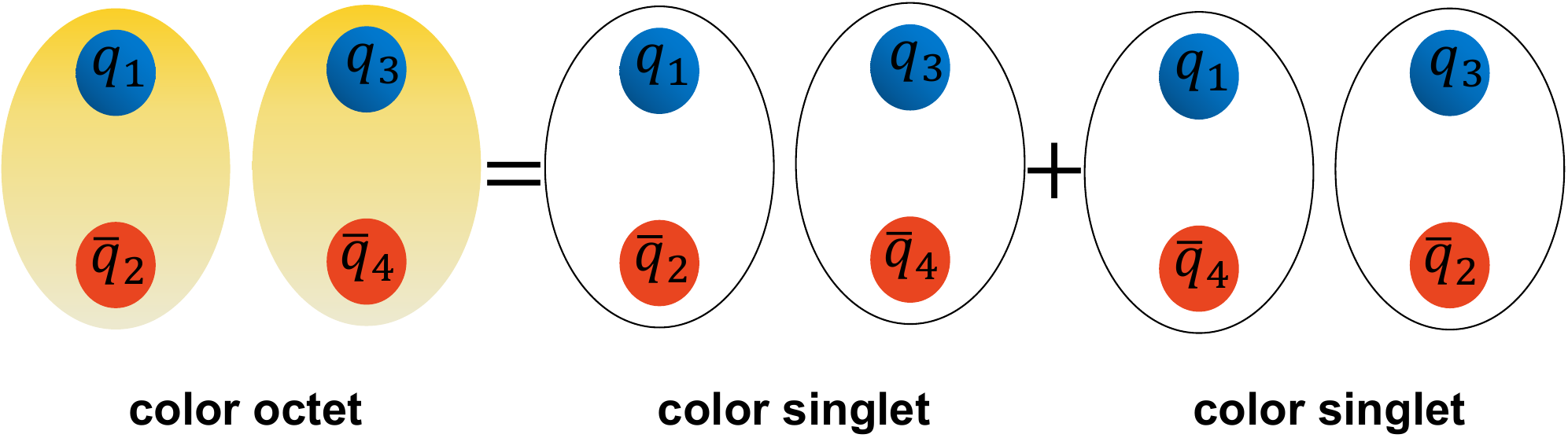}
  \caption{A color octet structure can be replaced by superimposing color singlets with the same quark order and another set of color singlets with a different quark order.}
   \label{MM1}
\end{figure}

Firstly, let's demonstrate that in the molecular structure, the contribution of the color octet state to the system can be substituted by a superposition of the color singlet ground state and excited states. Suppose there are four quarks in a four-quark system, two positive quarks are labeled as $q_1$ and $q_3$, and two anti-quarks are labeled as $\bar{q}_2$ and $\bar{q}_4$. Then, the three possible molecular states are as follows:$[q_1\bar{q}_2]_1 [q_3\bar{q}_4]_1$, $[q_1\bar{q}_2]_8 [q_3\bar{q}_4]_8$, and $[q_1\bar{q}_4]_1 [q_3\bar{q}_2]_1$. In the first case, ${q}_1$ and $\bar{q}_2$ form a color singlet, and ${q}_3$ and $\bar{q}_4$ form a color singlet. These two color singlets form a colorless $[q_1\bar{q}_2]_1 [q_3\bar{q}_4]_1$. In the second case, ${q}_1$ and $\bar{q}_2$ form a color octet, ${q}_3$ and $\bar{q}_4$ form a color octet, and these two color octets form a colorless $[q_1\bar{q}_2]_8 [q_3\bar{q}_4]_8$. In the third case, ${q}_1$ and $\bar{q}_4$ form a color singlet, and ${q}_3$ and $\bar{q}_2$ form a color singlet. These two color singlets form a colorless $[q_1\bar{q}_4]_1 [q_3\bar{q}_2]_1$. In the quark model, any physical wave function is composed of four parts, namely the spatial part, the spin part, the flavor part, and the color part. In molecular structures, the biggest difference among these three cases lies in the color wave functions. Based on the expressions of color singlet and color octet wave functions, it can be easily proven that $[q_1\bar{q}_2]_8 [q_3\bar{q}_4]_8$ can be composed of $[q_1\bar{q}_2]_1 [q_3\bar{q}_4]_1$ and $[q_1\bar{q}_4]_1 [q_3\bar{q}_2]_1$, which can be written as follows:
\begin{eqnarray*}
\label {q1}
[q_1\bar{q}_2]_8 [q_3\bar{q}_4]_8 = \frac{3}{\sqrt{8}}  [q_1\bar{q}_2]_1 [q_3\bar{q}_4]_1 - \frac{1}{\sqrt{8}} [q_1\bar{q}_4]_1 [q_3\bar{q}_2]_1
\end{eqnarray*}
Assuming that the relative motions between $q_1$-$\bar{q}_2$, $q_3$-$\bar{q}_4$, and $q_1\bar{q}_2$-$q_3\bar{q}_4$ are in their respective ground states in $[q_1\bar{q}_2]_8 [q_3\bar{q}_4]_8$, variations in quark order could lead to the three relative motions in $[q_1\bar{q}_4]_1 [q_3\bar{q}_2]_1$ being excited states. Thus, it is plausible to assert that the color octet state in a tetraquark structure can be substituted by a superposition of color singlet ground and excited states, which can be depicted by FIG. \ref{MM1},

Secondly,  by constructing complete superposition states of color singlet and color octet states, it can be proven that the diquark structure can be replaced by the dimeson structure. Here, we also define two diquark structures: $[q_1 q_3]_{\bar{3}} [\bar{q}_2\bar{q}_4]_3$, and $[q_1 q_3]_{6} [\bar{q}_2\bar{q}_4]_{\bar{6}}$. In the first case, ${q}_1$ and ${q}_3$ form an anti-color triplet, and $\bar{q}_2$ and $\bar{q}_4$ form a color triplet. Subsequently, the anti-color triplet and the color triplet are coupled into a colorless $[q_1q_3]_{\bar{3}} [\bar{q}_2\bar{q}_4]_3$. In the second case, ${q}_1$ and ${q}_3$ form the anti-color sextuple, and $\bar{q}_2$ and $\bar{q}_4$ form the color sextuple. Subsequently, the anti-color sextuple and the color sextuple are coupled into a colorless $[q_1q_3]_{6} [\bar{q}_2\bar{q}_4]_{\bar{6}}$. Likewise, according to the forms of the color wave functions for diquarks and dimesons, we can easily obtain the transformation relationship between their color wave functions.
\begin{align*}
[q_1q_3]_{\bar{3}} [\bar{q}_2\bar{q}_4]_3 = \sqrt{\frac{1}{3}}  [q_1\bar{q}_2]_1 [q_3\bar{q}_4]_1 - \sqrt{\frac{2}{3}} [q_1\bar{q}_2]_8 [q_3\bar{q}_4]_8, \\
[q_1q_3]_{6} [\bar{q}_2\bar{q}_4]_{\bar{6}} = \sqrt{\frac{2}{3}}  [q_1\bar{q}_2]_1 [q_3\bar{q}_4]_1 + \sqrt{\frac{1}{3}} [q_1\bar{q}_2]_8 [q_3\bar{q}_4]_8.
\end{align*}

As for the spatial part, we define the relative coordinates $\mathbf{r}_{ij}$ and $\mathbf{r}_{ij,kl}$ as follows:
\begin{eqnarray}
\begin{array}{cccccc}
\mathbf{r}_{ij}=\mathbf{r}_i-\mathbf{r}_j,~\mathbf{r}_{ij,kl}=\frac{m_i\mathbf{r}_i+m_j\mathbf{r}_j}{m_i+m_j}
-\frac{m_k\mathbf{r}_k+m_l\mathbf{r}_l}{m_k+m_l},\nonumber
\end{array}
\end{eqnarray}
where  \{$\mathbf{r}_{12}$, $\mathbf{r}_{34}$, $\mathbf{r}_{12,34}$\}  and \{$\mathbf{r}_{13}$, $\mathbf{r}_{24}$, $\mathbf{r}_{13,24}$\}  serve as the Jacobi coordinates for the dimeson configuration and the diquark configuration, respectively. Then, expressing the orbit wave function $\phi(\mathbf{r}_{12}, \mathbf{r}_{34},\mathbf{r}_{12,34})$ in terms of the explicit diquark configuration, denoted as $\phi(\mathbf{r}_{12},\mathbf{r}_{34},\mathbf{r}_{12,34})=\phi'(\mathbf{r}_{13},\mathbf{r}_{24},\mathbf{r}_{13,24})$. If we adopt the Gaussian function as the orbit trial wave function, the coordinate-related components can be expressed as

\begin{eqnarray}
&e^{-\nu_i\mathbf{r}^2_{12}-\nu_j\mathbf{r}^2_{34}-\nu_k\mathbf{r}^2_{12,34}}=e^{-a\mathbf{r}^2_{13}-b\mathbf{r}^2_{24}-c\mathbf{r}^2_{13,24}} \\
&\times e^{-d\mathbf{r}_{13}\cdot\mathbf{r}_{24}-e\mathbf{r}_{13}\cdot\mathbf{r}_{13,24}
-f\mathbf{r}_{24}\cdot\mathbf{r}_{13,24}},\nonumber
\end{eqnarray}

in which the coefficients $a$-$f$ depend on the variational parameters $\nu$s and the elements of the transformation matrix from \{$\mathbf{r}_{12}$, $\mathbf{r}_{34}$, $\mathbf{r}_{12,34}$\} to \{$\mathbf{r}_{13}$, $\mathbf{r}_{24}$, $\mathbf{r}_{13,24}$\}. The radial wave function includes all possible relative orbital angular momenta coupled to zero angular momentum~\cite{Vijande:2009ac},
\begin{widetext}
\begin{eqnarray}
&e^{-d\mathbf{r}_{13}\cdot\mathbf{r}_{24}-e\mathbf{r}_{13}\cdot\mathbf{r}_{13,24}
-f\mathbf{r}_{24}\cdot\mathbf{r}_{13,24}}=\frac{1}{4\sqrt{\pi}}\sum_{l_{13}=0}^{\infty}
\sum_{l_{24}=0}^{\infty}\sum_{l_{13,24}=0}^{\infty}
\left[[Y_{l_{13}}(\hat{\mathbf{r}}_{13})Y_{l_{24}}(\hat{\mathbf{r}}_{24})]_{l_{13,24}}
Y_{l_{13,24}}(\hat{\mathbf{r}}_{13,24})\right]_0\nonumber\\
&\times\sum_{l_1,l_2,l_3}\left(2l_1+1\right)\left(2l_2+1\right)\left(2l_3+1\right)\langle l_10l_20|l_{13}\rangle
\langle l_10l_30|l_{24}\rangle\langle l_20l_30|l_{13,24}\rangle\left \{
\begin{array}{cccccc}
l_{13} &  l_{24} &  l_{13,24} \\
\noalign{\smallskip}
l_3    &  l_2    &  l_1       \\
\end{array}
\right\}\nonumber\\
&\times\left(\sqrt{\frac{\pi}{2dr_{13}r_{24}}}I_{l_1+\frac{1}{2}}(dr_{13}r_{24})\right)
\left(\sqrt{\frac{\pi}{2er_{13}r_{13,24}}}I_{l_2+\frac{1}{2}}(er_{13}r_{13,24})\right)
\left(\sqrt{\frac{\pi}{2fr_{24}r_{13,24}}}I_{l_3+\frac{1}{2}}(fr_{24}r_{13,24})\right),\nonumber
\end{eqnarray}
\end{widetext}
where $I_a(x)$ are the modified Bessel functions.

\begin{figure}[h]
\vspace{-0.1cm}
\setlength {\abovecaptionskip} {+0.2cm}
\setlength {\belowcaptionskip} {+0.1cm}
  \centering
  \includegraphics[width=8cm,height=3cm]{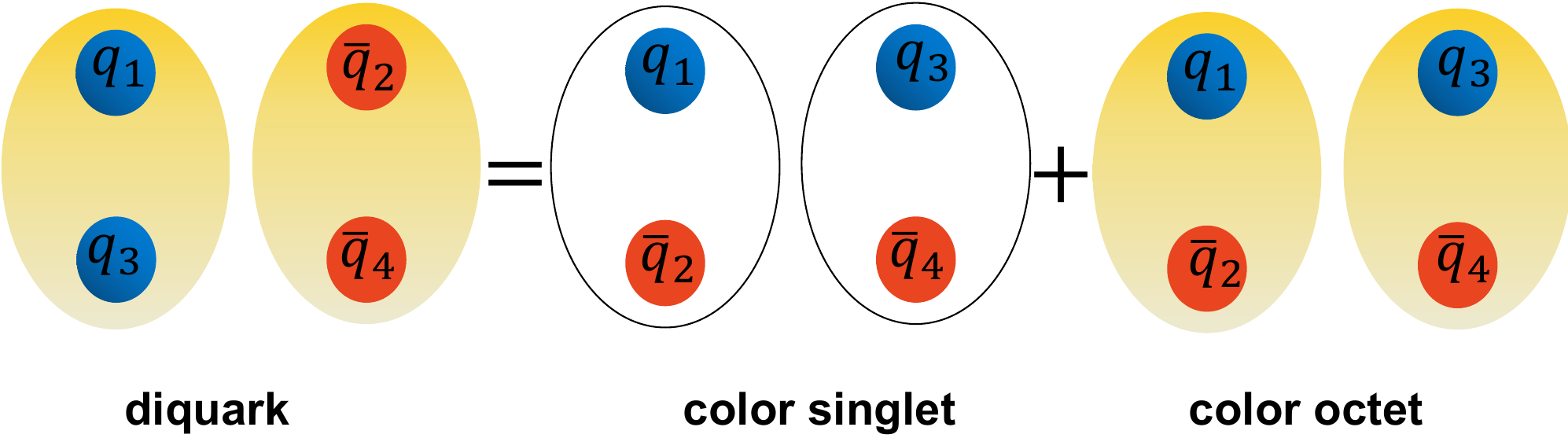}
  \caption{The diquark structure can be obtained by superimposing color singlets and color octets with identical quark content.}
   \label{DM2}
\end{figure}

Similar to the spatial part, the spin part also differs only between the dimeson structure and the diquark structure. Due to the constraints of quantum numbers, the total spin of the ground state $T_{cc}$ system is always $\mathbf{1}$. Thus, its spin of subgroups could have three possible combinations: \{$\mathbf{1}\oplus\mathbf{1}$, $\mathbf{1}\oplus\mathbf{0}$, $\mathbf{0}\oplus\mathbf{1}$\}. We denote the spin structures of dimesons and diquarks as \{$\mathbf{1}_{12}\oplus\mathbf{1}_{34}$, $\mathbf{1}_{12}\oplus\mathbf{0}_{34}$, $\mathbf{0}_{12}\oplus\mathbf{1}_{34}$\} and \{$\mathbf{1}_{13}\oplus\mathbf{1}_{24}$, $\mathbf{1}_{13}\oplus\mathbf{0}_{24}$, $\mathbf{0}_{13}\oplus\mathbf{1}_{24}$\}, respectively. The transformation relations between them are expressed as follows.
\begin{eqnarray}
\left (
\begin{array}{cccccc}
\mathbf{1}_{12}\oplus\mathbf{1}_{34}   \\
 \noalign{\smallskip}
\mathbf{1}_{12}\oplus\mathbf{0}_{34}    \\
\noalign{\smallskip}
\mathbf{0}_{12}\oplus\mathbf{1}_{34}    \\
\end{array}
\right )&=&\left (
\begin{array}{cccccc}
         0            & \frac{1}{\sqrt{2}} & \frac{1}{\sqrt{2}}   \\
\noalign{\smallskip}
-\frac{1}{\sqrt{2}}   &     \frac{1}{2}    & -\frac{1}{2}         \\
\noalign{\smallskip}
-\frac{1}{\sqrt{2}}   &     -\frac{1}{2}   &  \frac{1}{2}         \\
\end{array}
\right )\left (
\begin{array}{cccccc}
\mathbf{1}_{13}\oplus\mathbf{1}_{24}   \\
 \noalign{\smallskip}
\mathbf{1}_{13}\oplus\mathbf{0}_{24}    \\
\noalign{\smallskip}
\mathbf{0}_{13}\oplus\mathbf{1}_{24}    \\
\end{array}
\right ).\nonumber
\end{eqnarray}
Considering the $T_{cc}$ system, the flavor wave functions of dimesons and the flavor wave functions of diquark structures can also be mutually transformed. Therefore, from a theoretical perspective, we have demonstrated that if the wave functions are complete, then dimeson structures and diquark structures can be mutually transformed, which can be depicted by FIG. \ref{DM2},

In this paper, we will take the $T_{cc}$ system as an example, with the help of the multi-Gaussian expansion method. Two cases are studied systematically, namely, the equivalence between color octet and higher-order component wave, and the equivalence between diquark structure and molecular structure. The structure of this paper is organized as follows: In Section I, we provide an introduction to the study. In Section II, we elaborate on the details of the Chiral Quark Model (ChQM) and Gaussian Expansion Method (GEM) employed in our investigation. Subsequently, in Section III, we describe a method for identifying and calculating the decay width of the genuine resonance state. In Section IV, we present the numerical results. Finally, in Section IV, we summarize our findings and conclude this work.

\section{Chiral quark model, wave function of $T_{cc}$ system} \label{wavefunction and chiral quark model}
\subsection{Chiral quark model}
The chiral quark model has been applied successfully in describing the hadron spectra and hadron-hadron interactions.
The details of the model can be found in Refs. \cite{Vijande:2004he,Hu:2021nvs,Tan:2020ldi}. In this writing, we introduce a chiral quark model with scalar nonet exchange.
The Hamiltonian of the chiral quark model is given as follows,
\begin{eqnarray}
H &=& \sum_{i=1}^n(m_i+ \frac{{\vec{p}_i}^2}{2m_i})-T_{CM} +  \sum_{i<j=1}^n [ V(\textbf{r}_{ij}) ],
\end{eqnarray}
where $m_i$ is the constituent masse of $i$-th quark (antiquark), $\frac{\vec{p}_i^2}{2m_i}$ is the kinetic energy of $i$-th quark (antiquark),
$\textbf{r}_{ij}$ refers to the relative coordinates between $i$-th and $j$-th quark(antiquark).  $T_{CM}$ is centroid kinetic energy, which can be written as
\begin{eqnarray} \label{Tcm}
T_{CM}=\frac{(\vec{p}_{1}+\vec{p}_{2}+\vec{p}_{3}+\vec{p}_{4})^2}{2(m_1+m_2+m_3+m_4)}
\end{eqnarray}

The interaction potential $V(\textbf{r}_{ij})$,  shown in Eq.~\ref{Vr}, stems from the derivation of field theory and embodies three essential aspects of QCD: the confinement of quarks, the phenomenon of asymptotic freedom, and the spontaneous breaking of chiral symmetry.
\begin{eqnarray} \label{Vr}
  V(\textbf{r}_{ij})= V_{CON}(\textbf{r}_{ij}) + V_{OGE}(\textbf{r}_{ij}) + V_{OBE}(\textbf{r}_{ij})
\end{eqnarray}
The form of the quark confinement potential $V_{CON}(r_{ij})$ has not been rigorously proven. However, it is widely accepted that the exchange of multiple gluons results in a potential with a linearly increasing behavior. This potential is proportional to the distance between quarks, and this notion has been informally demonstrated in Ref.~\cite{Bali:2000gf}. The $V_{CON}(r_{ij})$ term includes central force $V_{con}^{C}(r_{ij})$ and spin-orbit force $V_{con}^{SO}(r_{ij})$.
\begin{subequations}
\label{Vcon}
\begin{align}
    V_{con}^C(r_{ij}) &= ( -a_{c} r_{ij}^{2}-\Delta) \boldsymbol{\lambda}_i^c \cdot \boldsymbol{\lambda}_j^c\\
    V_{con}^{SO}(r_{ij}) &= -\boldsymbol{\lambda}_i^c \cdot \boldsymbol{\lambda}_j^c \frac{a_c}{4m_i^2m_j^2}[ ((m_i^2+m_j^2)(1-2a_s) \nonumber \\
&+ 4m_im_j(1-a_s))(\vec{S}_{+}\cdot \vec{L})+ \nonumber \\
&((m_j^2-m_i^2)(1-2a_s))(\vec{S}_{-}\cdot \vec{L}) ]
\end{align}
\end{subequations}
Due to the absence of a rigorous proof for the form of the confinement potential, there are currently three prevalent models: linear confinement potential \cite{Yang:2020atz}, quadratic confinement potential, and confinement potential with screening effects \cite{Vijande:2004he}. In this paper, we adopt the quadratic confinement potential. In Eqs.~\ref{Vcon}, the parameters $a_c$, $a_s$, and $\Delta$ are defined and listed in Table \ref{modelparameters}, and $\vec{S}_{\pm}=\vec{S}_{i}\pm\vec{S}_j$. $\boldsymbol{\lambda}^{c}$ are $SU(3)$ color Gell-Mann matrices.

\begin{table}[t]
\begin{center}
\caption{Quark model parameters ($m_{\pi}=0.7$ $fm^{-1}$, $m_{\sigma}=3.42$ $fm^{-1}$, $m_{\eta}=2.77$ $fm^{-1}$).\label{modelparameters}}
\begin{tabular}{cccc}
\hline\hline\noalign{\smallskip}
Quark masses   &$m_u=m_d$(MeV)       &313  \\
               &$m_{c}$(MeV)         &1728 \\
\hline
Goldstone bosons
                   &$\Lambda_{\pi}(fm^{-1})$     &4.2  \\
                   &$\Lambda_{\eta}(fm^{-1})$    &5.2  \\
                   &$\Lambda_{a0}(fm^{-1})$      &4.2  \\
                   &$\Lambda_{f0}(fm^{-1})$      &5.8  \\
                   &$g_{ch}^2/(4\pi)$            &0.54 \\
                   &$\theta_p(^\circ)$           &-15  \\
\hline
Confinement        &$a_{c}$(MeV$\cdot fm^{-2}$)                 &101  \\
                   &$\Delta$(MeV)                &-78.3\\
\hline
OGE                &$\alpha_{qq}$                &0.5723\\
                   &$\alpha_{qc}$                &0.4938\\
                   &$\alpha_{cc}$                &0.3753\\
                   &$\hat{r}_0$(MeV)             &28.17 \\
                   &$\hat{r}_g$(MeV)             &34.5  \\
                   &$a_{s}$                      &0.777 \\
\hline\hline
\end{tabular}
\end{center}
\end{table}

Beyond the scale of chiral symmetry breaking, one expects that the dynamics are influenced by QCD perturbative effects. These effects mimic gluon fluctuations around the instanton vacuum and are incorporated through the $V_{OGE}(\textbf{r}_{ij})$ which contains central force $V_{oge}^{C}(r_{ij})$, spin-orbit force $V_{oge}^{SO}(r_{ij})$ and tensor force $V_{oge}^{T}(r_{ij})$. The central potential can be written as
\begin{eqnarray}
V_{oge}^C(r_{ij}) =\frac{\alpha_s}{4} \boldsymbol{\lambda}_i^c \cdot \boldsymbol{\lambda}_{j}^c \left[\frac{1}{r_{ij}}-\frac{2\pi}{3} \frac{\boldsymbol{\sigma}_i\cdot \boldsymbol{\sigma}_j}{m_im_j} \delta(r_{ij}) \right],
\end{eqnarray}
where $\boldsymbol{\sigma}$ are the $SU(2)$ Pauli matrices, $r_{0}(\mu_{ij})=\frac{r_0}{\mu_{ij}}$ and $\alpha_{s}$ is an effective scale-dependent running coupling,
\begin{equation}
 \alpha_s(\mu_{ij})=\frac{\alpha_0}{\ln\left[(\mu_{ij}^2+\mu_0^2)/\Lambda_0^2\right]}.
\end{equation}
The $\delta(r_{ij})$ function, arising as a consequence of the non-relativistic reduction of the one-gluon exchange diagram between point-like particles, has to be regularized in order to perform exact calculations. It reads
\begin{eqnarray}
\delta{(\boldsymbol{r}_{ij})}=\frac{e^{-r_{ij}/r_0(\mu_{ij})}}{4\pi r_{ij}r_0^2(\mu_{ij})}. \nonumber
\end{eqnarray}
The spin-orbit force $V_{oge}^{SO}(r_{ij})$ and tensor force $V_{oge}^{T}(r_{ij})$ can be written as
\begin{subequations}
\label{Voge}
\begin{align}
    V_{oge}^{SO}(r_{ij}) &= -\frac{1}{16} \frac{\alpha_s\boldsymbol{\lambda}_i^c \cdot \boldsymbol{\lambda}_j^c}{4m_i^2m_j^2}[\frac{1}{r_{ij}^3}-\frac{e^{-r_{ij}/r_g(\mu)}}{r_{ij}^3}(1+\frac{r_{ij}}{r_g(\mu)})] \nonumber \\
    &[ (m_i^2+m_j^2+4m_im_j)(\vec{S}_{+}\cdot \vec{L})\nonumber \\
    &+(m_j^2-m_i^2)(\vec{S}_{-}\cdot \vec{L}) ],\\ \nonumber
V_{oge}^{T}(r_{ij}) &= -\frac{1}{16} \frac{\alpha_s\boldsymbol{\lambda}_i^c \cdot \boldsymbol{\lambda}_j^c}{4m_i^2m_j^2}[\frac{1}{r_{ij}^3}-\frac{e^{-r_{ij}/r_g(\mu)}}{r_{ij}}(\frac{1}{r_{ij}^2} \nonumber \\
&+\frac{1}{3r^2_g(\mu)} +\frac{1}{r_{ij}r_g(\mu)})]S_{ij},
\end{align}
\end{subequations}
where $S_{ij}=3(\vec{\sigma}_i\cdot\hat{r}_{ij})(\vec{\sigma}_j\cdot\hat{r}_{ij})-(\vec{\sigma}_i\cdot\vec{\sigma}_j)$, $r_{g}(\mu_{ij})=\frac{r_g}{\mu_{ij}}$.

In the chiral quark model, quarks, referred to as constituent quarks, acquire their masses from the interaction with nearly massless current quarks and the gluon medium. To ensure the chiral invariance of QCD, there should be the exchange of Goldstone bosons between constituent quarks, giving rise to new Lagrangian interaction terms. Therefore, this interaction can be expressed as
\begin{eqnarray}
V_{OBE}(\textbf{r}_{ij})=V_{\chi}(\textbf{r}_{ij})+V_{S}(\textbf{r}_{ij}).
\end{eqnarray}
There are two different contributions including pseudoscalar meson exchange $V_{\chi=\pi,\eta,K}(\textbf{r}_{ij})$ and scalar meson exchange $V_{S=a_0,f_0,\kappa,\sigma}(\textbf{r}_{ij})$ to the Goldstone boson exchange. The pseudoscalar meson exchange potential $V_{\chi}(\textbf{r}_{ij})$ consists of two components: the central potential $v_{\chi=\pi,K,\eta}^C$ and the tensor force $v_{{\chi}=\pi,K,\eta}^T$. Simultaneously, the scalar meson exchange potential $V_{S}(\textbf{r}_{ij})$ encompasses the central potential $v_{s=\sigma,a_0,f_0,\kappa}^C$, along with the spin-orbit coupling potential $v_{s=\sigma,a_0,f_0,\kappa}^{SO}$.

\begin{align}
\begin{split}
 \left \{
\begin{array}{ll}
V_{\chi}(r_{ij}) & =  v_{\pi}({{\bf r}_{ij}})\sum_{a=1}^{3} \boldsymbol{\lambda}_i^a \boldsymbol{\lambda}_j^a+v_{K}({{\bf r}_{ij}})\sum_{a=4}^{7}
	\boldsymbol{\lambda}_i^a \boldsymbol{\lambda}_j^a \\
&+v_{\eta}({{\bf r}_{ij}})[\cos\theta_{P}(\boldsymbol{\lambda}_i^8 \boldsymbol{\lambda}_j^8)-\sin\theta_{P}(\boldsymbol{\lambda}_i^0 \boldsymbol{\lambda}_j^0)] ,  \\
v_{\chi=\pi,K,\eta}^C & =  \frac{g^2_{ch}}{4\pi} \frac{m_{\chi}^2}{\Lambda_{\chi}^2-m_{\chi}^2} \frac{\Lambda^2_{\chi}}{\Lambda^2_{\chi}-m^2_{\chi}}m_{\chi}[ Y(m_{{\chi}}r_{ij})-\frac{\Lambda_{\chi}^3}{m_{\chi}^3}\\
&Y(\Lambda_{{\chi}}r_{ij})](\vec{\sigma}_i\cdot\vec{\sigma}_j),   \\
v_{{\chi}=\pi,K,\eta}^T & =  \frac{g^2_{ch}}{4\pi} \frac{m_{{\chi}}^2}{\Lambda_{\chi}^2-m_{\chi}^2} \frac{\Lambda^2_{\chi}}{\Lambda^2_{\chi}-m^2_{\chi}}m_{\chi}
	[ H(m_{{\chi}}r_{ij})-\frac{\Lambda_{\chi}^3}{m_{\chi}^3}\\
&H(\Lambda_{{\chi}}r_{ij})]S_{ij},   \\
\end{array}
\right.
\end{split}
\end{align}

\begin{align}
\begin{split}
 \left \{
\begin{array}{ll}
V_{S}(r_{ij}) & =  v_{\sigma}({{\bf r}_{ij}}) \boldsymbol{\lambda}_i^0 \boldsymbol{\lambda}_j^0+v_{a_0}({{\bf r}_{ij}})\sum_{a=1}^{3} \boldsymbol{\lambda}_i^a \boldsymbol{\lambda}_j^a  \\
&  +v_{\kappa}({{\bf r}_{ij}})\sum_{a=4}^{7}
	\boldsymbol{\lambda}_i^a \boldsymbol{\lambda}_j^a+v_{f_0}({{\bf r}_{ij}})
    \boldsymbol{\lambda}_i^8 \boldsymbol{\lambda}_j^8, \\
v_{s=\sigma,a_0,f_0,\kappa}^C & =  -\frac{g^2_{ch}}{4\pi} \frac{\Lambda^2_s}{\Lambda^2_s-m^2_s}m_s
	\left[ Y(m_{s}r_{ij})-\frac{\Lambda_s}{m_s}Y(\Lambda_{s}r_{ij})\right],   \\
v_{s=\sigma,a_0,f_0,\kappa}^{SO} & =  -\frac{g^2_{ch}}{4\pi} \frac{\Lambda^2_s}{\Lambda^2_s-m^2_s}\frac{m_s^3}{2m_im_j}
	[ G(m_{s}r_{ij}) \\
&-\frac{\Lambda_s^3}{m_s^3}G(\Lambda_{s}r_{ij})]\vec{L}\cdot \vec{S},
\end{array}
\right.
\end{split}
\end{align}

where $\boldsymbol{\lambda}^a$ are $SU(3)$ flavor Gell-Mann matrices, $m_{\chi(s)}$ is the masses of Goldstone bosons, $\Lambda_{\chi(s)}$ is the cut-offs, $g^2_{ch}/4\pi$ is the Goldstone-quark coupling constant. Finally, $Y(x)$ is the standard Yukawa function defined by $Y(x)=e^{-x}/x$, $G(x)=(1+1/x)Y(x)/x$ and $H(x)=(1+3/x+3/x^2)Y(x)/x$.

All the parameters are determined by fitting the meson spectrum, taking into account only a quark-antiquark component. They are shown in Table~\ref{modelparameters}. The calculated masses of the mesons involved in the present work are shown in Table~\ref{mesonmass}.
\begin{table}[]
\caption{ \label{mesonmass}  Numerical results for the meson spectrum (in MeV) for different models. (unit: MeV).\label{mesons}}
\begin{tabular}{ccccccc}
\hline\hline\noalign{\smallskip}
    Meson           &    Our work     &GM\cite{Godfrey:2015dva} & ChQM\cite{Vijande:2004he} &  RM\cite{Ni:2021pce} & EXP.(PDG)             \\ \hline
    $D$             &    1892.34         &     1877                &  1883                      &   1865               & 1864.84$\pm$0.05      \\
    $D^*$           &    1980.37         &     2041                &  2010                      &   2008               & 2006.85$\pm$0.05      \\
    $D_0^*$         &    2332.70         &     2399                &    -                       &   2313               & 2343$\pm$10           \\
    $D_1^*$         &    2436.03         &     2456                &  2492                      &   2424               & $2420.1\pm0.6$        \\
    $D_2^*$         &    2472.99         &     2502                &  2502                      &   2475               & $2461.1^{+0.7}_{-0.8}$\\
    $D_1^{\prime}$  &    2448.54         &     2467                &    -                       &   2453               & $2412\pm9$            \\
\hline\hline
\end{tabular}
\end{table}

\subsection{The wave function of $T_{cc}$ system}
The wave function of $T_{cc}$ four quark system includes two important structures, namely dimeson structure and diquark structure. The wave function of each structure includes four parts, namely orbital, spin, flavor and color. The wave function of each part is constructed in two steps, first write down the two-body wave functions, Then coupling two sub-clusters wave functions to form the four-body one.

In this paper, we employ the $J$-$J$ coupling scheme to derive our wave functions. In the initial stage, we couple the orbital wave functions $\phi_{nlm}$ and spin wave functions $\chi_{S,mS}$ of the two subgroups to obtain the wave function $\psi_{J,mJ}(\textbf{r})$, which can be written as
\begin{eqnarray} \label{J1J2}
\psi_{J_1,mJ_1}(\boldsymbol{\lambda})= \phi_{n_1L_1m_1}(\boldsymbol{\lambda}) \otimes \chi_{S_1,mS_1}, \\
\psi_{J_2,mJ_2}(\boldsymbol{\rho})   = \phi_{n_2L_2m_2}(\boldsymbol{\rho}) \otimes \chi_{S_2,mS_2}.
\end{eqnarray}
Subsequently, the wave functions of these two subgroups, $\psi_{J_1,mJ_1}(\textbf{r})$ and $\psi_{J_2,mJ_2}(\textbf{r})$, are further coupled to form $\psi_{J_{12},mJ_{12}}(\textbf{r})$. This obtained wave function is then combined with the relative motion wave function $\phi_{n_3l_3m_3}(\hat{R})$ between the two subgroups to obtain the orbit-spin wave function $\psi^{SO}_{i}(\textbf{r})$, where $J$ denotes the total angular momentum.
\begin{eqnarray} \label{J1J2}
\psi_{J_{12},mJ_{12}}= \psi_{J_1,mJ_1}(\boldsymbol{\lambda}) \otimes \psi_{J_2,mJ_2}(\boldsymbol{\rho}), \\
\psi^{SO}_{i}(\textbf{r} )= \psi_{J_{12},mJ_{12}}\otimes \phi_{n_3L_3m_3}(\textbf{R}),\\
i \equiv \{L_1,S_1,J_1,L_2,S_2,J_2,J_{12},L_3 \}.
\end{eqnarray}

\begin{table}[]
\caption{ \label{JJ}  Different combinations of J-J coupling.\label{mesons}}
\begin{tabular}{cccccccc}
\hline\hline\noalign{\smallskip}
\multicolumn{2}{l}{$J_{12}=0,L_3=1$} & \multicolumn{2}{c}{$J_{1}=1$} &~~~~&\multicolumn{2}{c}{$J_{2}=1$}& i   \\
                                     && $L_1=0$ & $S_1=1$            & & $L_2=1$ & $S_2=0$           & 1       \\
                                     && $L_1=0$ & $S_1=1$            & & $L_2=1$ & $S_2=1$           & 2       \\
                                     \cline{3-7}
                                     && \multicolumn{2}{c}{$J_{1}=2$}& &\multicolumn{2}{c}{$J_{2}=2$}& i   \\
                                     && $L_1=2$ & $S_1=0$             && $L_2=1$ & $S_2=0$           & 3       \\
                                     && $L_1=2$ & $S_1=1$             && $L_2=1$ & $S_2=1$           & 4       \\ \hline
\multicolumn{2}{l}{$J_{12}=1,L_3=0$} & \multicolumn{2}{c}{$J_{1}=0$} &&\multicolumn{2}{c}{$J_{2}=1$}& i   \\
                                     && $L_1=0$ & $S_1=0$             && $L_2=0$ & $S_2=1$           & 5       \\
                                     && $L_1=1$ & $S_1=1$             && $L_2=1$ & $S_2=0$           & 6       \\
                                     && $L_1=1$ & $S_1=1$             && $L_2=1$ & $S_2=1$           & 7       \\
                                     \cline{3-7}
                                     && \multicolumn{2}{c}{$J_{1}=1$} &&\multicolumn{2}{c}{$J_{2}=1$}& i   \\
                                     && $L_1=0$ & $S_1=1$             && $L_2=0$ & $S_2=1$           & 8       \\
                                     && $L_1=0$ & $S_1=1$             && $L_2=2$ & $S_2=1$           & 9       \\
                                     && $L_1=1$ & $S_1=0$             && $L_2=1$ & $S_2=0$           & 10       \\
                                     && $L_1=1$ & $S_1=0$             && $L_2=1$ & $S_2=1$           & 11       \\
                                     && $L_1=1$ & $S_1=1$             && $L_2=1$ & $S_2=1$           & 12       \\
                                     && $L_1=2$ & $S_1=1$             && $L_2=2$ & $S_2=1$           & 13      \\ \hline
\multicolumn{2}{l}{$J_{12}=1,L_3=2$} & \multicolumn{2}{c}{$J_{1}=0$} &&\multicolumn{2}{c}{$J_{2}=1$}& i   \\
                                     && $L_1=1$ & $S_1=1$             && $L_2=1$ & $S_2=0$           & 14       \\
                                     && $L_1=0$ & $S_1=1$             && $L_2=1$ & $S_2=1$           & 15       \\
                                     && $L_1=0$ & $S_1=0$             && $L_2=0$ & $S_2=1$           & 16      \\
                                     && $L_1=1$ & $S_1=0$             && $L_2=1$ & $S_2=0$           & 17       \\
                                     \cline{3-7}
                                     && \multicolumn{2}{c}{$J_{1}=1$} &&\multicolumn{2}{c}{$J_{2}=1$}& i   \\
                                     && $L_1=1$ & $S_1=0$             && $L_2=1$ & $S_2=0$           & 18       \\
                                     && $L_1=1$ & $S_1=0$             && $L_2=1$ & $S_2=1$           & 19       \\
                                     && $L_1=1$ & $S_1=1$             && $L_2=1$ & $S_2=1$           & 20       \\
                                     && $L_1=0$ & $S_1=1$             && $L_2=0$ & $S_2=1$           & 21       \\
                                     && \multicolumn{2}{c}{$J_{1}=2$} &&\multicolumn{2}{c}{$J_{2}=1$}& i   \\
                                     && $L_1=2$ & $S_1=0$             && $L_2=0$ & $S_2=1$           & 22       \\
                                     && $L_1=2$ & $S_1=1$             && $L_2=0$ & $S_2=1$           & 23       \\
                                     && $L_1=1$ & $S_1=1$             && $L_2=1$ & $S_2=0$           & 24       \\
                                     && $L_1=1$ & $S_1=1$             && $L_2=1$ & $S_2=1$           & 25       \\
                                     && \multicolumn{2}{c}{$J_{1}=2$} &&\multicolumn{2}{c}{$J_{2}=2$}& i   \\
                                     && $L_1=1$ & $S_1=1$             && $L_2=1$ & $S_2=1$           & 27       \\
\hline
\end{tabular}
\end{table}

In this coupling process, there are a total of 8 variables, denoted as $L_1,S_1,J_1,L_2,S_2,J_2,J_{12},L_3$. For simplicity, we use ``i" to represent their combinations. Considering the symmetry of $T_{cc}$, such combinations  are considered in 27 cases, all listed in Table~\ref{JJ}. Because of no difference between spin of quark and antiquark, the meson-meson structure has the same spin wave function as the diquark-antidiquark structure. The spin wave functions of the sub-cluster are shown below,
\begin{align*}
&\chi^{11}_{\sigma}=\alpha\alpha,~~
\chi^{10}_{\sigma}=\frac{1}{\sqrt{2}}(\alpha\beta+\beta\alpha),~~
\chi^{1-1}_{\sigma}=\beta\beta,\nonumber \\
&\chi^{00}_{\sigma}=\frac{1}{\sqrt{2}}(\alpha\beta-\beta\alpha).
\end{align*}
In the context of quark spin, $\alpha$ and $\beta$ represent the third component of quark spin, taking values of $\frac{1}{2}$ and $-\frac{1}{2}$, respectively, in two distinct cases.

The spatial wave functions are Gaussians with range parameters chosen to lie in a geometrical progression:
\begin{subequations}
 \label{orbit}
 \begin{align}
\phi_{n_1L_1m_1}(\boldsymbol{\lambda})&=N_{n_1L_1}r^{L_1}e^{-(\lambda/\lambda_{n1})^2}Y_{L_1m_1}(\hat{\lambda}), \\
\phi_{n_2L_2m_2}(\boldsymbol{\rho})   &=N_{n_2L_2}r^{L_2}e^{-(   \rho/\rho_{n2})^2}Y_{L_2m_2}(\hat{\rho}), \\
\phi_{n_3L_3m_3}(\boldsymbol{R})      &=N_{n_3L_3}r^{L_3}e^{-(      R/R_{n3})^2}Y_{L_3m_3}(\hat{R}),
 \end{align}
\end{subequations}
 where $N_{n_1L_1},N_{n_2L_2}$ and $N_{n_3L_3}$ are normalization constants, which can be expressed by a general formula
\begin{align}
\label{Nnl}
N_{nL}=\left[ \frac{2^{L+2}(2\nu_n)^{L+\frac{3}{2}}}{\sqrt{\pi}(2L+1)!!} \right]^{\frac{1}{2}}.
\end{align}
In this Eq.~\ref{Nnl}, $\nu_n$ represents the expansion width determined by the range parameter $r_n$ of the wave function, which can be expressed as $\nu_n=\frac{1}{r_n^2}$. For the motion between different quarks, the expression for the range parameter is as follows:
\begin{subequations}
 \label{rn}
 \begin{align}
\lambda_{n_1}&=\lambda_{min}a_1^{n_1-1}, a_1={\frac{\lambda_{max}}{\lambda_{min}}}^{\frac{1}{n_1-1}},\\
\rho_{n_2}&=\rho_{min}a_2^{n_2-1}, a_2={\frac{\rho_{max}}{\rho_{min}}}^{\frac{1}{n_2-1}}, \\
R_{n_3}&=R_{min}a_3^{n_3-1}, a_3={\frac{R_{max}}{R_{min}}}^{\frac{1}{n_3-1}},
 \end{align}
\end{subequations}
 where, $n_1$, $n_2$, and $n_3$ represent the Gaussian expansion orders for the three relative motions among the four quarks. $\lambda_{min}$ and $\lambda_{max}$ denote the minimum and maximum values for the Gaussian expansion of the first subgroup's relative motion. Similarly, $\rho_{min}$ and $\rho_{max}$ represent the minimum and maximum values for the Gaussian expansion of the second subgroup's relative motion. $R_{min}$ and $R_{max}$ indicate the minimum and maximum values for the Gaussian expansion between these two groups. The range parameters here are all obtained through fitting the meson spectrum.
\begin{figure*}[htbp]
\centering

\subfigure[ ]{
\begin{minipage}[t]{0.5\linewidth}
\centering
\includegraphics[width=4cm,height=3cm]{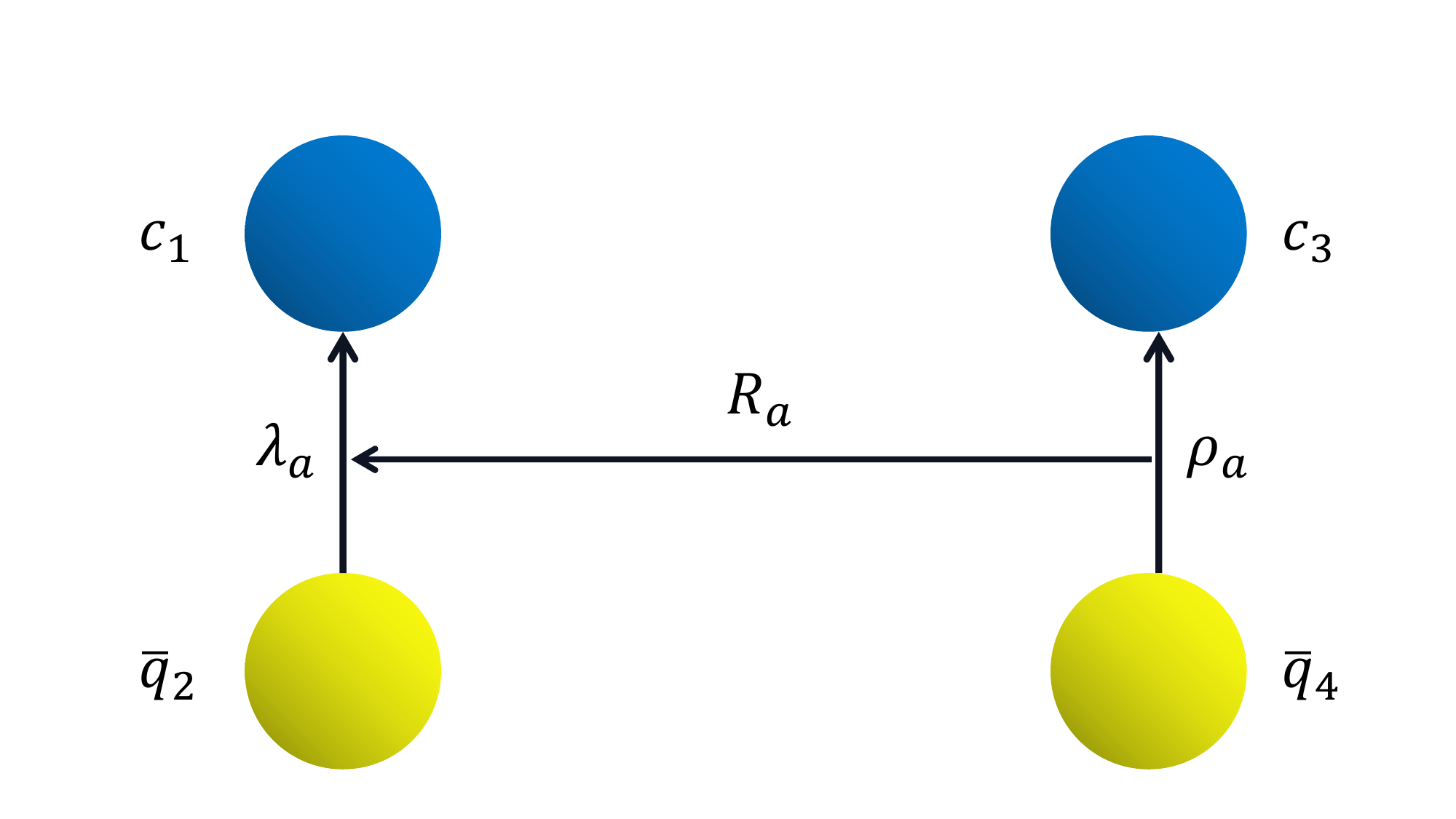}
\end{minipage}%
}%
\subfigure[ ]{
\begin{minipage}[t]{0.5\linewidth}
\centering
\includegraphics[width=4cm,height=3cm]{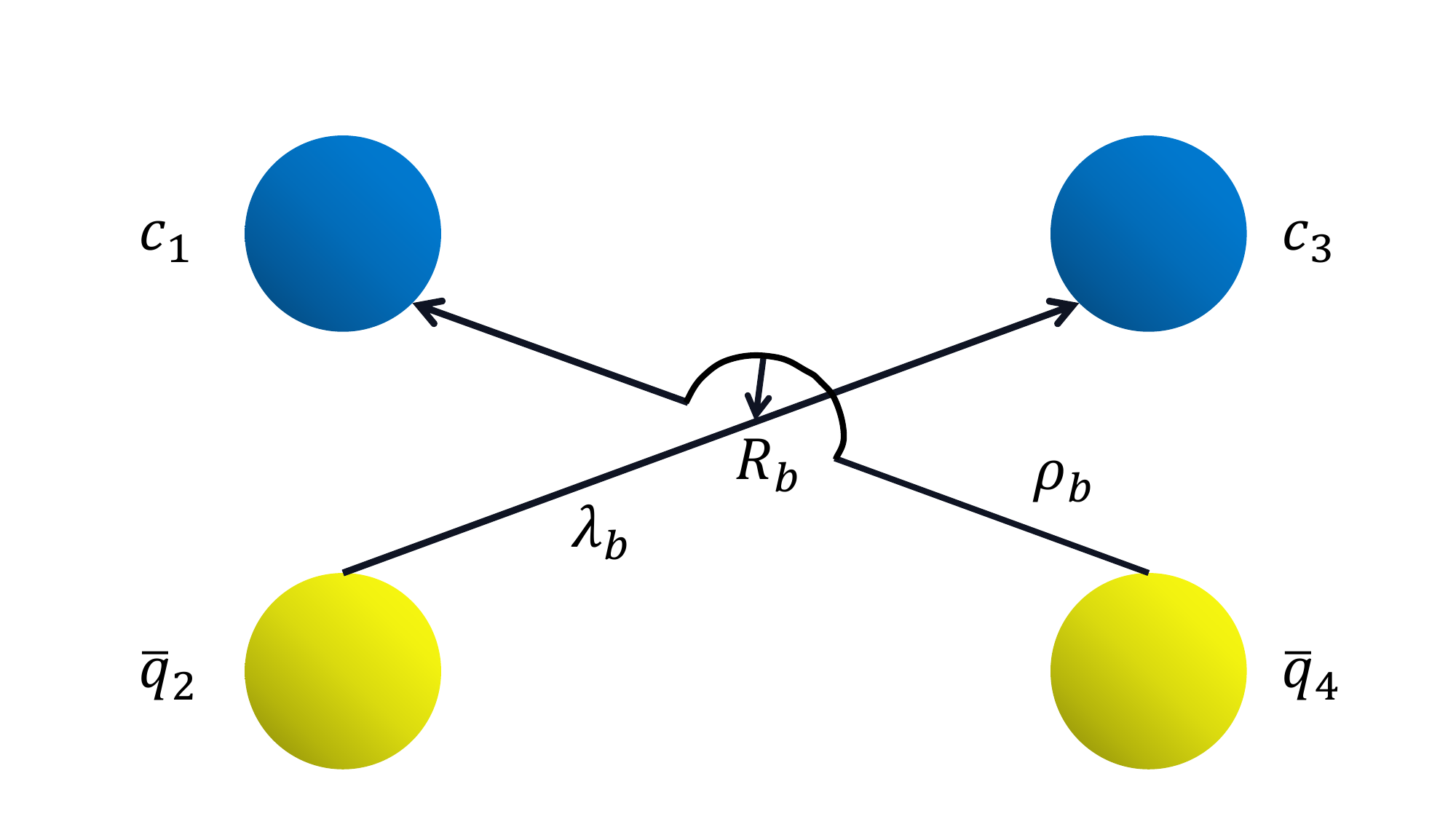}
\end{minipage}%
}
\subfigure[ ]{
\begin{minipage}[t]{0.3\linewidth}
\centering
\includegraphics[width=4cm,height=3cm]{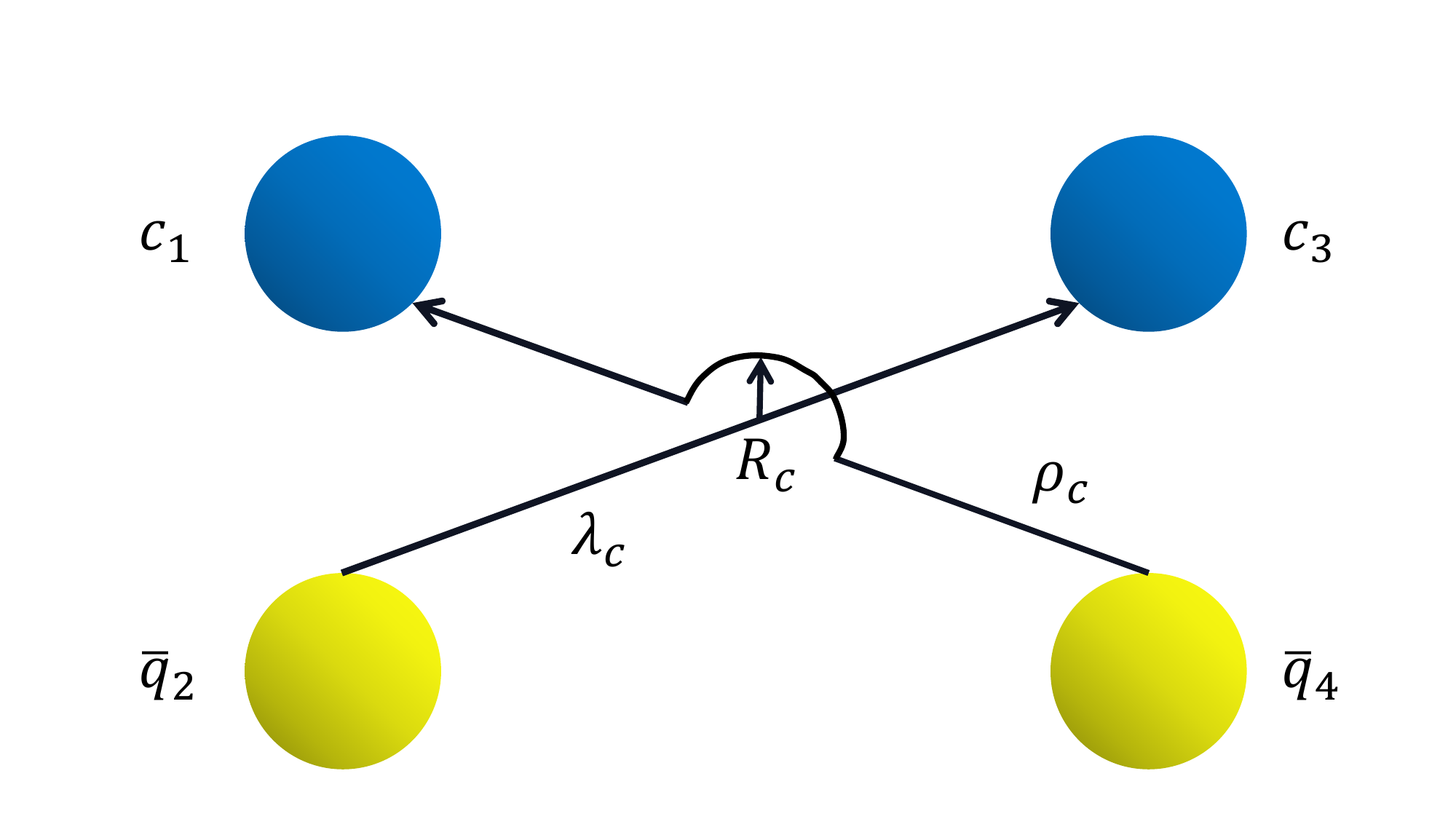}
\end{minipage}
}
\subfigure[ ]{
\begin{minipage}[t]{0.3\linewidth}
\centering
\includegraphics[width=4cm,height=3cm]{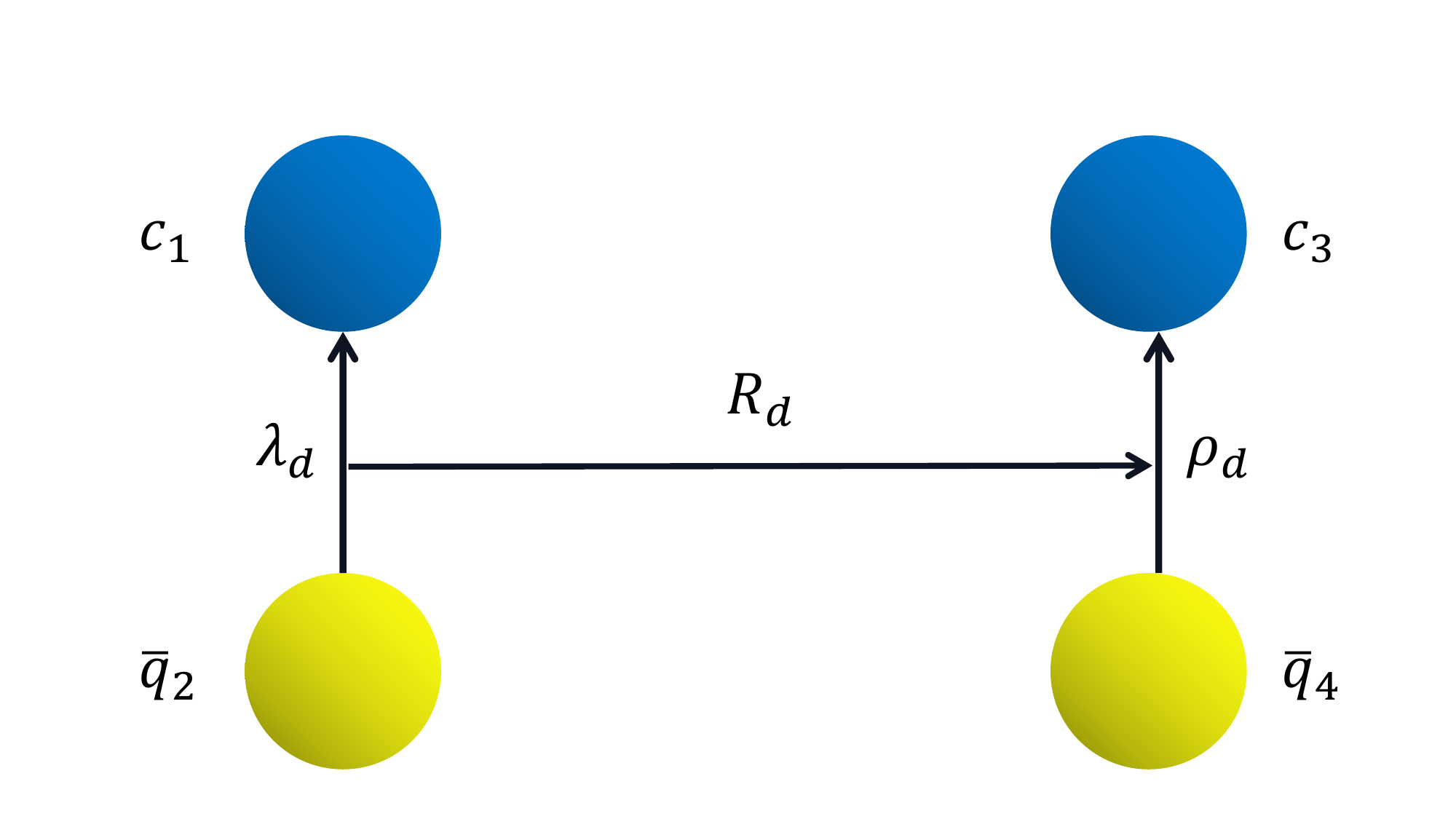}
\end{minipage}
}
\subfigure[ ]{
\begin{minipage}[t]{0.3\linewidth}
\centering
\includegraphics[width=4cm,height=3cm]{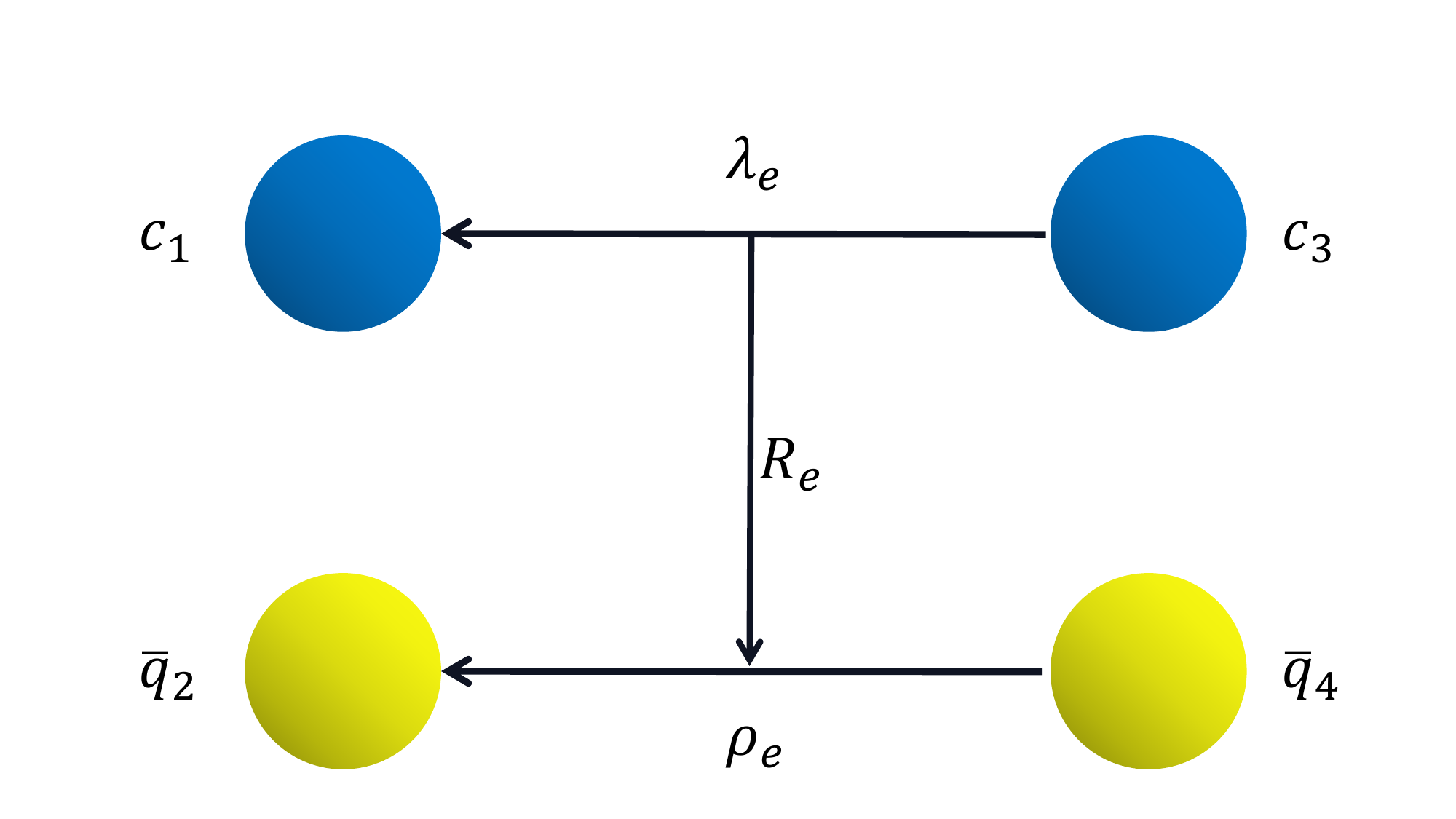}
\end{minipage}
}

\centering
\caption{ The spatial structures of the dimeson are represented (a), (b), (c) and (d),  while (e) represents the spatial structure of the diquark wave function.}.\label{Fic}
\end{figure*}
 We have two flavor wave functions of the $T_{cc}$ system,
\begin{align*}
\xi_{1}=(c\bar{q})(c\bar{q}),\xi_{2}=(cc)(\bar{q}\bar{q}).
\end{align*}
$\xi_{1}$ is for meson-meson structure, and $\xi_{2}$ is for diquark-antidiquark structure.
The colorless tetraquark system has four color wave functions, two for meson-meson structure, $1\otimes1$ ($\varphi_{1}$), $8\otimes8$ ($\varphi_{2}$),
and two for diquark-antidiquark structure, $\bar{3}\otimes 3$ ($\varphi_{3}$) and $6\otimes \bar{6}$ ($\varphi_{4}$).
\begin{eqnarray}
\varphi_{1} &= & \sqrt{\frac{1}{9}}(\bar{r}r\bar{r}r+\bar{r}r\bar{g}g+\bar{r}r\bar{b}b+\bar{g}g\bar{r}r+\bar{g}g\bar{g}g \nonumber\\
 & & +\bar{g}g\bar{b}b+\bar{b}b\bar{r}r+\bar{b}b\bar{g}g+\bar{b}b\bar{b}b), \nonumber \\
\varphi_{2} & = & \sqrt{\frac{1}{72}}(3\bar{b}r\bar{r}b+3\bar{g}r\bar{r}g+3\bar{b}g\bar{g}b+3\bar{g}b\bar{b}g+3\bar{r}g\bar{g}r \nonumber \\
& & +3\bar{r}b\bar{b}r+2\bar{r}r\bar{r}r+2\bar{g}g\bar{g}g+2\bar{b}b\bar{b}b-\bar{r}r\bar{g}g \nonumber \\
& & -\bar{g}g\bar{r}r-\bar{b}b\bar{g}g-\bar{b}b\bar{r}r-\bar{g}g\bar{b}b-\bar{r}r\bar{b}b), \nonumber \\
\varphi_{3} &= &
 \sqrt{\frac{1}{12}}(rg\bar{r}\bar{g}-rg\bar{g}\bar{r}+gr\bar{g}\bar{r}-gr\bar{r}\bar{g}+rb\bar{r}\bar{b} \nonumber \\
 & & -rb\bar{b}\bar{r}+br\bar{b}\bar{r}-br\bar{r}\bar{b}+gb\bar{g}\bar{b}-gb\bar{b}\bar{g} \nonumber \\
 & & +bg\bar{b}\bar{g}-bg\bar{g}\bar{b}), \nonumber \\
\varphi_{4} &= & \sqrt{\frac{1}{24}}(2rr\bar{r}\bar{r}+2gg\bar{g}\bar{g}+2bb\bar{b}\bar{b}
    +rg\bar{r}\bar{g}+rg\bar{g}\bar{r} \nonumber \\
& & +gr\bar{g}\bar{r}+gr\bar{r}\bar{g}+rb\bar{r}\bar{b}+rb\bar{b}\bar{r}+br\bar{b}\bar{r} \nonumber \\
& & +br\bar{r}\bar{b}+gb\bar{g}\bar{b}+gb\bar{b}\bar{g}+bg\bar{b}\bar{g}+bg\bar{g}\bar{b}).
\end{eqnarray}

\begin{table*}[th]
  \centering
  \fontsize{12}{9}\selectfont
  \makebox[\textwidth][c]{
   \begin{threeparttable}
   \caption{The masses of $D_s$ mesons and the percentages of $c\bar{s}$ component in the states. `S', `P' and `D' in the parentheses denote
   $S$-, $P$- and $D$-wave states.}\label{try}
    \begin{tabular}{ccccccccc}
    \hline\hline
 Index & $|i,j,k\rangle$ & color singlet: $J_{12}=0,L_3=1$       &~~~~~&Index &$|i,j,k\rangle$&color singlet:$J_{12}=1,L_3=2$         \\ \hline
1&$|1,1,1\rangle$&$D^{*}D_{1}^{'}$-$P$~wave                            &&12&$|14,1,1\rangle$&$D_0^{*}D_{1}^{'}$-$D$~wave              \\
2&$|2,1,1\rangle$&$D^{*}D_{1}^{*}(P)$-$P$~wave                         &&13&$|15,1,1\rangle$&$D_0^{*}D_{1}^{*}(P)$-$D$~wave           \\
3&$|3,1,1\rangle$&$D_2^{'}D_{2}^{*}(P)$-$P$~wave                       &&14&$|16,1,1\rangle$&$DD^{*}$-$D$~wave                        \\
4&$|4,1,1\rangle$&$D_2^{*}(D)D_2^{*}(P)$-$P$~wave                      &&15&$|17,1,1\rangle$&$D_1^{'}D_{1}^{'}$-$D$~wave              \\ \cline{1-3}
Index &$|i,j,k\rangle$ &color singlet: $J_{12}=1,L_3=0$                &&16&$|18,1,1\rangle$&$D_1^{'}D_{1}^{*}(P)$-$D$~wave           \\ \cline{1-3}
 &$|5,1,1\rangle$&$DD^{*}$-$S$~wave                                    &&17&$|19,1,1\rangle$&$D_1^{*}(P)D_{1}^{*}(P)$-$D$~wave        \\
5&$|6,1,1\rangle$&$D_0^{*}D_1^{'}(P)$-$S$~wave                         &&18&$|20,1,1\rangle$&$D^{*}D^{*}$-$D$~wave                    \\
6&$|7,1,1\rangle$&$D_0^{*}D_1^{*}(P)$-$S$~wave                         &&19&$|21,1,1\rangle$&$D_2^{'} D^{*}$-$D$~wave                 \\
 &$|8,1,1\rangle$&$D^{*}D^{*}$-$S$~wave                                &&20&$|22,1,1\rangle$&$D_2^{*}(D)D^{*}$-$D$~wave               \\
7&$|9,1,1\rangle$&$D^{*}D_1^{*}(D)$-$S$~wave                           &&21&$|23,1,1\rangle$&$D_2^{*}(P)D_1^{'}$-$D$~wave             \\
8&$|10,1,1\rangle$&$D_1^{'}D_1^{'}$-$S$~wave                           &&22&$|24,1,1\rangle$&$D_2^{*}(P)D_{1}^{*}(P)$-$D$~wave        \\
9&$|11,1,1\rangle$&$D_1^{'}D_1^{*}(P)$-$S$~wave                        &&23&$|25,1,1\rangle$&$D_2^{*}(P)D_{2}^{*}(P)$-$D$~wave        \\
10&$|12,1,1\rangle$&$D_1^{*}(P)D_1^{*}(P)$-$S$~wave                                                                                   \\
11&$|13,1,1\rangle$&$D_1^{*}(D)D_1^{*}(D)$-$S$~wave                                                                                   \\ \hline
 Index & $|i,j,k\rangle$ & color octet: $J_{12}=0,L_3=1$         &~~~~~&Index &$|i,j,k\rangle$&color octet:$J_{12}=1,L_3=2$                           \\ \hline
24&$|1,1,2\rangle$&$[D^{*}D_{1}^{'}]_8$-$P$~wave                       &&37&$|14,1,2\rangle$&$[D_0^{*}D_{1}^{'}]_8$-$D$~wave          \\
25&$|2,1,2\rangle$&$[D^{*}D_{1}^{*}(P)]_8$-$P$~wave                    &&38&$|15,1,2\rangle$&$[D_0^{*}D_{1}^{*}(P)]_8$-$D$~wave       \\
26&$|3,1,2\rangle$&$[D_2^{'}D_{2}^{*}(P)]_8$-$P$~wave                  &&39&$|16,1,2\rangle$&$[DD^{*}]_8$-$D$~wave                    \\
27&$|4,1,2\rangle$&~~~~~$[D_2^{*}(D)D_2^{*}(P)]_8$-$P$~wave            &&40&$|17,1,2\rangle$&$[D_1^{'}D_{1}^{'}]_8$-$D$~wave          \\ \cline{1-3}
Index &$|i,j,k\rangle$ &color octet: $J_{12}=1,L_3=0$                  &&41&$|18,1,2\rangle$&$[D_1^{'}D_{1}^{*}(P)]_8$-$D$~wave       \\ \cline{1-3}
28&$|5,1,2\rangle$&$[DD^{*}]_8$-$S$~wave                               &&42&$|19,1,2\rangle$&$[D_1^{*}(P)D_{1}^{*}(P)]_8$-$D$~wave    \\
29&$|6,1,2\rangle$&$[D_0^{*}D_1^{'}(P)]_8$-$S$~wave                    &&43&$|20,1,2\rangle$&$[D^{*}D^{*}]_8$-$D$~wave                \\
30&$|7,1,2\rangle$&$[D_0^{*}D_1^{*}(P)]_8$-$S$~wave                    &&44&$|21,1,2\rangle$&$[D_2^{'} D^{*}]_8$-$D$~wave             \\
31&$|8,1,2\rangle$&$[D^{*}D^{*}]_8$-$S$~wave                           &&45&$|22,1,2\rangle$&$[D_2^{*}(D)D^{*}]_8$-$D$~wave           \\
32&$|9,1,2\rangle$&$[D^{*}D_1^{*}(D)]_8$-$S$~wave                      &&46&$|23,1,2\rangle$&$[D_2^{*}(P)D_1^{'}]_8$-$D$~wave         \\
33&$|10,1,2\rangle$&$[D_1^{'}D_1^{'}]_8$-$S$~wave                      &&47&$|24,1,2\rangle$&$[D_2^{*}(P)D_{1}^{*}(P)]_8$-$D$~wave    \\
34&$|11,1,2\rangle$&$[D_1^{'}D_1^{*}(P)]_8$-$S$~wave                   &&48&$|25,1,2\rangle$&~~~~$[D_2^{*}(P)D_{2}^{*}(P)]_8$-$D$~wave\\
35&$|12,1,2\rangle$&$[D_1^{*}(P)D_1^{*}(P)]_8$-$S$~wave      \\
36&$|13,1,2\rangle$&$[D_1^{*}(D)D_1^{*}(D)]_8$-$S$~wave      \\    \hline
 Index & $|i,j,k\rangle$ & diquark: $J_{12}=0,L_3=1$         \\    \hline
49&$|5,2,4\rangle$ & $[cc]_6^0[\bar{q}\bar{q}]_{\bar{6}}^1$-$S$~wave  \\
50&$|8,2,3\rangle$ & $[cc]_{\bar{3}}^1[\bar{q}\bar{q}]_{3}^1$-$S$~wave\\
\hline
    \end{tabular}
   \end{threeparttable}}
  \end{table*}

The symbols $r (\bar{r})$, $g (\bar{g})$, and $b (\bar{b})$ correspond to the colors of quarks (antiquarks), representing red, green, and blue, respectively.

Based on the above discussion, the $T_{cc}$ tetraquark system is composed of three parts: the spin-orbit coupled wave function $\psi^{SO}_{i}(\textbf{r} )$, the flavor wave function $\xi_{j}$, and the color wave function $\varphi_{k}$. We have listed all possible physical channels in Table \ref{try}. The total wave function $\Psi_{i,j,k}(\textbf{r})$ can be written as  below

\begin{align}
 \label{phi}
  \Psi_{i,j,k}(\textbf{r}) &= \mathcal{A} \psi^{SO}_{i}(\textbf{r}) \xi_{j} \varphi_{k}
 \end{align}

The total wave function $\Psi_{i,j,k}(\textbf{r})$ can be expanded as follows:

 \begin{align}
 \label{Y_Ex}
  &\Psi_{i,j,k}(\textbf{r}) = C_{J_{12},mJ_{12};L_3,mL_3}^{J,mJ}C_{J_1,mJ_{1};J_2,mJ_2}^{J_{12},mJ_{12}}C_{L_1,mL_{1};S_1,mS_1}^{J_1,mJ_1} \nonumber\\
  &C_{L_2,mL_{2};S_2,mS_2}^{J_2,mJ_2} \mathcal{A} \phi_{n_1L_1m_1}(\boldsymbol{\lambda})\phi_{n_2L_2m_2}(\boldsymbol{\rho})\phi_{n_3L_3m_3}(\boldsymbol{R}) \nonumber \\
  &\chi_{S_1,mS_1}\chi_{S_2,mS_2} \xi_{j} \varphi_{k}
 \end{align}

In Eq.~\ref{Y_Ex}, $C$ represents the Clebsch-Gordan coefficient. $\mathcal{A}$ is the antisymmetrization operator, and for the dimeson structure in the $T_{cc}$ system, ${\cal A}=1-P_{13}-P_{24}+P_{13}P_{24}$, where $P_{ij}$ denotes the antisymmetric operator, and the four exchange terms correspond to the first four quark coupling orders in Fig.~\ref{Fic}. However, for the diquark structure in the $T_{cc}$ system, due to symmetry considerations, ${\cal A}=1$, corresponding to the fifth quark coupling order in Fig.~\ref{Fic}. The relevant coordinates in each set are defined as

 \begin{align}
  \boldsymbol{\lambda}_{a}&=\textbf{r}_1-\textbf{r}_2,\boldsymbol{\rho}_{a}=\textbf{r}_3-\textbf{r}_4, \nonumber \\
  \textbf{R}_{a}&=\frac{m_1 \textbf{r}_1 +m_2 \textbf{r}_2}{m_1+m_2}-\frac{m_3 \textbf{r}_3 +m_4 \textbf{r}_4}{m_3+m_4}, \\
  \boldsymbol{\lambda}_{b}&=\textbf{r}_1-\textbf{r}_4,\boldsymbol{\rho}_{b}=\textbf{r}_2-\textbf{r}_3, \nonumber \\
  \textbf{R}_{b}&=\frac{m_1 \textbf{r}_1 +m_4 \textbf{r}_4}{m_1+m_4}-\frac{m_2 \textbf{r}_2 +m_3 \textbf{r}_3}{m_2+m_3}, \\
  \boldsymbol{\lambda}_{c}&=\textbf{r}_3-\textbf{r}_2,\boldsymbol{\rho}_{c}=\textbf{r}_1-\textbf{r}_4, \nonumber \\
  \textbf{R}_{c}&=\frac{m_3 \textbf{r}_3 +m_2 \textbf{r}_2}{m_3+m_2}-\frac{m_1 \textbf{r}_1 +m_4 \textbf{r}_4}{m_1+m_4},
 \end{align}
 \begin{align}
  \boldsymbol{\lambda}_{d}&=\textbf{r}_3-\textbf{r}_4,\boldsymbol{\rho}_{d}=\textbf{r}_1-\textbf{r}_2, \nonumber \\
  \textbf{R}_{d}&=\frac{m_3 \textbf{r}_4 +m_4 \textbf{r}_4}{m_3+m_4}-\frac{m_1 \textbf{r}_1 +m_2 \textbf{r}_2}{m_1+m_2}, \\
  \boldsymbol{\lambda}_{e}&=\textbf{r}_1-\textbf{r}_3,\boldsymbol{\rho}_{e}=\textbf{r}_2-\textbf{r}_4, \nonumber \\
  \textbf{R}_{e}&=\frac{m_1 \textbf{r}_1 +m_3 \textbf{r}_3}{m_1+m_3}-\frac{m_2 \textbf{r}_2 +m_4 \textbf{r}_4}{m_2+m_4}.
 \end{align}
In all cases, the center-of-mass coordinate is
\begin{equation}
\textbf{R}=\frac{m_1 \textbf{r}_1+m_2 \textbf{r}_2+m_3 \textbf{r}_3+m_4 \textbf{r}_4}{m_1+m_2+m_3+m_4},
\end{equation}
where $m_1$, $m_2$, $m_3$, and $m_4$ are the masses of particles 1, 2, 3, and 4, respectively.

At last, we solve the following Schr\"{o}dinger equation to obtain eigen-energies of the system, with the help of the Rayleigh-Ritz variational principle.
\begin{equation}
H \Psi_{i,j,k}(\textbf{r}) =E \Psi_{i,j,k}(\textbf{r}),
\end{equation}
the eigenvalues ($E_n$) and eigenvectors ($C_n$) of the system are obtained by solving the diagonalization problem,
 \begin{align}
[(H)-E_n(N)][C_n]=0,
 \end{align}
with
\begin{subequations}
 \label{Hpsi}
 \begin{align*}
(H)=\langle \Psi_{i,j,k}(\textbf{r})|H|\Psi_{i,j,k}(\textbf{r}) \rangle,  \\
(N)=\langle \Psi_{i,j,k}(\textbf{r})|1|\Psi_{i,j,k}(\textbf{r}) \rangle.
 \end{align*}
\end{subequations}

In the quark model, obtaining matrix elements that involve tensor forces and spin-orbit coupling is a complex process, especially when dealing with wave functions associated with orbital excited states. Taking the structure in Fig.~\ref{Fic}(a) as an example, where the orbital quantum number between $c_1$ and $\bar{q}_2$ is 1, after Jacobi coordinate transformation to the structure in Fig.~\ref{Fic}(b), where $c_1$ forms a $c_1$-$\bar{q}_4$ subgroup with $\bar{q}_4$, the orbital quantum number between $c_1$ and $\bar{q}_4$ cannot be directly determined. In this paper, we employ the spherical harmonics expansion method (Eq.~(139) in Ref.~\cite{Hiyama:2003cu}). If there is the following relationship between Jacobi coordinates: $\bf{r_{a}}=\alpha \bf{r_{b}}+\beta \bf{R_{b}}$, then the transformation relationship between their spherical harmonics is given by

 \begin{align}
 \label{Y_Ex}
  &r_a^{L} Y_{L,mL}(\textbf{r}_{a}) = \sum\limits _{\lambda=0}^{L} \left[   \frac{4 \pi (2L+1) !}{(2 k +1)! ( 2(L-k)!+1)!} \right]^{1/2}(\alpha r_b)^{k} \nonumber\\
  &(\beta R_b)^{L-k} \sum\limits _{mk=-k}^{k}  C_{k,mk;L-k,mL-mk}^{L,mL} Y_{k,mk}(\textbf{r}_b) \nonumber \\
  & Y_{L-k,mL- mk}(\textbf{R}_b).
 \end{align}

In Eq.~\ref{Y_Ex}, $k$ represents the possible orbital angular momentum of $\textbf{r}_b$, $L-k$ denotes the orbital angular momentum of $\textbf{R}_b$, and their sum corresponds to the orbital angular momentum of $\textbf{r}_a$. The subscripts $mk$ and $mL-mk$ refer to the respective third components. In the $T_{cc}$ system, we need to handle matrix elements like $\langle \Psi_{i,j,k}(\textbf{r}_a)|V(r_c)|\Psi_{i,j,k}(\textbf{r}_b) \rangle$. This involves Jacobi coordinate transformations, i.e., transforming $\Psi_{i,j,k}(\textbf{r}_a)$ and $\Psi_{i,j,k}(\textbf{r}_b)$ into the form of $\Psi_{i,j,k}(\textbf{r}_c)$, and then considering spatial integrations. We first assume that there is the following Jacobi coordinate transformation relationship between the Jacobi coordinates of $r_a$ and $r_c$,
\begin{eqnarray}
\lambda_a= \alpha_1 \lambda_{c1} +\beta_1 \rho_{c2} +\gamma_1 R_{c3}, \\
   \rho_a= \alpha_2 \lambda_{c1} +\beta_2 \rho_{c2} +\gamma_2 R_{c3}, \\
      R_a= \alpha_3 \lambda_{c1} +\beta_3 \rho_{c2} +\gamma_3 R_{c3},
\end{eqnarray}
where, $\alpha$, $\beta$, and $\gamma$ are transformation coefficients for the coordinates $\textbf{r}_a$ to transform into $\textbf{r}_c$. Then, $\lambda_a^{L_{a1}} Y_{L_{a1},mL_{a1}}(\boldsymbol{\lambda}_{a})$,$\rho_a^{L_{a2}} Y_{L_{a2},mL_{a2}}(\boldsymbol{\rho}_{a})$,$R_a^{L_{a3}} Y_{L_{a3},mL_{a3}}(\boldsymbol{R}_a)$ can be expanded as:
\begin{widetext}
 \begin{align}
 \label{Rab}
  \lambda_a^{L_{a1}} Y_{L_{a1},mL_{a1}}(\boldsymbol{\lambda}_{a1})& = \sum\limits _{k_1=0}^{L_{a1}} \sum\limits _{t_1=0}^{k_1} \left[   \frac{4 \pi (2L_{a1}+1) !}{(2 k_1 +1)! ( 2(L_{a1}-k_1)!+1)!} \right]^{1/2} \left[   \frac{4 \pi (2 k_1+1) !}{(2 t_1 +1)! ( 2(k_1-t_1)!+1)!} \right]^{1/2} (\alpha_1 \lambda_{c1})^{t_1} (\beta_1 \rho_{c2})^{k_1-t_1}  \nonumber  \\
  &(\gamma_1 R_{c3})^{L_{a1}-k_1}  \sum\limits _{mk_1=-k_1}^{k_{1}} \sum\limits _{mt_1=-t_1}^{t_1} C_{k_1,mk_1;L_{a1}-k_1,mL_{a1}-mk_1}^{L_{a1},mL_{a1}} C_{t_1,mt_1;k_1-t_1,mk_1-mt_1}^{k_1,mt_1}   Y_{t_1,mt_1}(\boldsymbol{\lambda}_{c1})   \nonumber \\
  &Y_{k_1-t_1,mk_1- mt_1}(\boldsymbol{\rho}_{c2})Y_{L_{a1}-k_1,mL_{a1}- mk_1}(\textbf{R}_{c3}). \nonumber \\
   \rho_a^{L_{a1}} Y_{L_{a2},mL_{a2}}(\boldsymbol{\rho}_{a2})& = \sum\limits _{k_2=0}^{L_{a2}} \sum\limits _{t_2=0}^{k_2} \left[   \frac{4 \pi (2L_{a2}+1) !}{(2 k_2 +1)! ( 2(L_{a2}-k_2)!+1)!} \right]^{1/2} \left[   \frac{4 \pi (2 k_2+1) !}{(2 t_2 +1)! ( 2(k_2-t_2)!+1)!} \right]^{1/2} (\alpha_2 \lambda_{c1})^{t_2} (\beta_2 \rho_{c2})^{k_2-t_2}  \nonumber  \\
  &(\gamma_2 R_{c3})^{L_{a2}-k_2}  \sum\limits _{mk_2=-k_2}^{k_{2}} \sum\limits _{mt_2=-t_2}^{t_2} C_{k_2,mk_2;L_{a2}-k_2,mL_{a2}-mk_2}^{L_{a2},mL_{a2}} C_{t_2,mt_2;k_2-t_2,mk_2-mt_2}^{k_2,mt_2}   Y_{t_2,mt_2}(\boldsymbol{\lambda}_{c1})   \nonumber \\
  &Y_{k_2-t_2,mk_2- mt_2}(\boldsymbol{\rho}_{c2})Y_{L_{a2}-k_2,mL_{a2}- mk_2}(\textbf{R}_{c3}). \nonumber \\
   R_a^{L_{a3}} Y_{L_{a3},mL_{a3}}(\boldsymbol{R}_{a3})& = \sum\limits _{k_3=0}^{L_{a3}} \sum\limits _{t_3=0}^{k_3} \left[   \frac{4 \pi (2L_{a3}+1) !}{(2 k_3 +1)! ( 2(L_{a3}-k_3)!+1)!} \right]^{1/2} \left[   \frac{4 \pi (2 k_3+1) !}{(2 t_3 +1)! ( 2(k_3-t_3)!+1)!} \right]^{1/2} (\alpha_3 \lambda_{c1})^{t_3} (\beta_3 \rho_{c2})^{k_3-t_3}  \nonumber  \\
  &(\gamma_3 R_{c3})^{L_{a3}-k_3}  \sum\limits _{mk_3=-k_3}^{k_{3}} \sum\limits _{mt_3=-t_3}^{t_3} C_{k_3,mk_3;L_{a3}-k_3,mL_{a3}-mk_3}^{L_{a3},mL_{a3}} C_{t_3,mt_3;k_3-t_3,mk_3-mt_3}^{k_3,mt_3} Y_{t_3,mt_3}(\boldsymbol{\lambda}_{c1})   \nonumber \\
  &Y_{k_3-t_3,mk_3- mt_3}(\boldsymbol{\rho}_{c2})Y_{L_{a3}-k_3,mL_{a3}- mk_3}(\textbf{R}_{c3}),
 \end{align}
\end{widetext}
where the total wave function $\Psi_{i,j,k}(\textbf{r}_a)$ for the three relative motions is expanded into nine relative motion wave functions. Within these nine functions, there are actually three different orbital quantum numbers denoted as $Y_{L_1,m_1}(\boldsymbol{\lambda}_{c1})$, $Y_{L_2,m_2}(\boldsymbol{\rho}_{c2})$, and $Y_{L_3,m_3}(\boldsymbol{R}_{c3})$. Next, we need to combine the spherical harmonics with different orbital quantum numbers using the following formula:
 \begin{align}
 \label{rr}
 Y_{L_1,m_1}(\textbf{r})Y_{L_2,m_2}(\textbf{r}) =\sum\limits _{L=|L_1-L_2|}^{L_1+L_2}C_{L_1,m_1;L_2,m_2}^{L,m_1+m_2}C_{L_1,0;L_2,0}^{L,0} \nonumber \\
 \left[ \frac{(2L_1+1)(2L_2+1)}{4\pi(2L+1)} \right]^{1/2} Y_{L,m_1+m_2}(\textbf{r})
 \end{align}

Therefore, the total wave function $\Psi_{i,j,k}(\textbf{r}_a)$ undergoes a transformation based on Jacobi coordinate change (Eqs.~\ref{Jacobi}). Initially, the spatial wave function for $r_a$ transforms from three terms to nine terms, introducing six additional CG coefficients according to Eq.~\ref{Y_Ex}, when transforming from the initial $r_a$ to any coordinate $r_c$. The spherical harmonics with the same orbital quantum numbers in the transformed wave function are then combined using Eq.~\ref{rr}. The final total wave function is expressed as follows:
\begin{widetext}
 \begin{align}
 \label{Y_Ex}
  \Psi_{i,j,k}(\textbf{r}) = &C_{J_{12},mJ_{12};L_3,mL_3}^{J,mJ}C_{J_1,mJ_{1};J_2,mJ_2}^{J_{12},mJ_{12}}C_{L_1,mL_{1};S_1,mS_1}^{J_1,mJ_1}C_{L_2,mL_{2};S_2,mS_2}^{J_2,mJ_2}\chi_{S_1,mS_1}\chi_{S_2,mS_2} \xi_{j} \varphi_{k}\sum\limits _{k_1=0}^{L_{a1}} \sum\limits _{t_1=0}^{k_1}  \sum\limits _{k_2=0}^{L_{a2}} \sum\limits _{t_2=0}^{k_2} \sum\limits _{k_3=0}^{L_{a3}} \sum\limits _{t_3=0}^{k_3} \nonumber \\
  &\left[   \frac{4 \pi (2L_{a1}+1) !}{(2 k_1 +1)! ( 2(L_{a1}-k_1)!+1)!} \right]^{1/2} \left[   \frac{4 \pi (2 k_1+1) !}{(2 t_1 +1)! ( 2(k_1-t_1)!+1)!} \right]^{1/2} \left[   \frac{4 \pi (2L_{a2}+1) !}{(2 k_2 +1)! ( 2(L_{a2}-k_2)!+1)!} \right]^{1/2}   \nonumber  \\
  & \left[   \frac{4 \pi (2 k_2+1) !}{(2 t_2 +1)! ( 2(k_2-t_2)!+1)!} \right]^{1/2} \left[   \frac{4 \pi (2L_{a3}+1) !}{(2 k_3 +1)! ( 2(L_{a3}-k_3)!+1)!} \right]^{1/2} \left[   \frac{4 \pi (2 k_3+1) !}{(2 t_3 +1)! ( 2(k_3-t_3)!+1)!} \right]^{1/2} \nonumber  \\
  &(\alpha_1 \lambda_{c1})^{t_1} (\beta_1 \rho_{c2})^{k_1-t_1} (\gamma_1 R_{c3})^{L_{a1}-k_1} (\alpha_2 \lambda_{c1})^{t_2} (\beta_2 \rho_{c2})^{k_2-t_2} (\gamma_2 R_{c3})^{L_{a2}-k_2}(\alpha_3 \lambda_{c1})^{t_3} (\beta_3 \rho_{c2})^{k_3-t_3}(\gamma_3 R_{c3})^{L_{a3}-k_3} \nonumber  \\
  & \sum\limits _{mk_1=-k_1}^{k_{1}} \sum\limits _{mt_1=-t_1}^{t_1} \sum\limits _{mk_2=-k_2}^{k_{2}} \sum\limits _{mt_2=-t_2}^{t_2} \sum\limits _{mk_3=-k_3}^{k_{3}} \sum\limits _{mt_3=-t_3}^{t_3}
  C_{k_1,mk_1;L_{a1}-k_1,mL_{a1}-mk_1}^{L_{a1},mL_{a1}} C_{t_1,mt_1;k_1-t_1,mk_1-mt_1}^{k_1,mt_1} \nonumber  \\
  &C_{k_2,mk_2;L_{a2}-k_2,mL_{a2}-mk_2}^{L_{a2},mL_{a2}} C_{t_2,mt_2;k_2-t_2,mk_2-mt_2}^{k_2,mt_2}C_{k_3,mk_3;L_{a3}-k_3,mL_{a3}-mk_3}^{L_{a3},mL_{a3}}C_{t_3,mt_3;k_3-t_3,mk_3-mt_3}^{k_3,mt_3}
  \nonumber \\
  &\sum\limits _{t_{12}=|t_1-t_2|}^{t_1+t_2} \sum\limits _{t_{123}=|t_{12}-t_3|}^{t_{12}+t_3}\sum\limits _{s_{12}=|(k_1-t_1)-(k_2-t_2)|}^{|(k_1-t_1)+(k_2-t_2)|} \sum\limits _{s_{123}=|s_{12}-(k_3-t_3)|}^{s_{12}+(k_3-t_3)}\sum\limits _{q_{12}=|(L_{a1}-k_1)-(L_{a2}-k_2)|}^{|(L_{a1}-k_1)+(L_{a2}-k_2)|} \sum\limits _{q_{123}=|q_{12}-(L_{a3}-k_3)|}^{q_{12}+(L_{a3}-k_3)} \nonumber  \\
 &\left[ \frac{(2t_1+1)(2t_2+1)}{4\pi(2t_{12}+1)} \right]^{1/2}\left[ \frac{(2t_{12}+1)(2t_3+1)}{4\pi(2t_{123}+1)} \right]^{1/2} \left[ \frac{(2(k_1-t_1)+1)(2(k_2-t_2)+1)}{4\pi(2s_{12}+1)} \right]^{1/2} \nonumber  \\
 &\left[ \frac{(2s_{12}+1)(2(k_3-t_3)+1)}{4\pi(2s_{123}+1)} \right]^{1/2} \left[ \frac{(2(L_{a1}-k_1)+1)(2(L_{a2}-k_2)+1)}{4\pi(2q_{12}+1)} \right]^{1/2} \left[ \frac{(2q_{12}+1)(2(L_{a3}-k_3)+1)}{4\pi(2q_{123}+1)} \right]^{1/2}  \nonumber  \\
 & C_{t_1,mt_1;t_2,mt_2}^{t_{12},mt_{12}}C_{t_1,0;t_2,0}^{t_{12},0}C_{t_{12},mt_{12};t_3,mt_3}^{t_{123},mt_{123}}C_{t_{12},0;t_3,0}^{t_{123},0} C_{(L_{a1}-k_1),(mL_{a1}-mk_1);(L_{a2}-k_2),(mL_{a2}-mk_2)}^{q_{12},mq_{12}}C_{(L_{a1}-k_1),0;(L_{a2}-k_2),0}^{q_{12},0}  \nonumber  \\ &C_{q_{12},mq_{12};(L_{a3}-k_3),(mL_{a3}-mk_3)}^{q_{123},mq_{123}}C_{q_{12},0;(L_{a3}-k_3),0}^{q_{123},0} C_{(k_{1}-t_1),(mk_1-mt_1);(k_2-t_2),(mk_2-mt_2)}^{s_{12},ms_{12}}C_{(k_{1}-t_1),0;(k_2-t_2),0}^{s_{12},0} \nonumber  \\
 &C_{s_{12},ms_{12};(k_3-t_3),(mk_3-mt_3)}^{s_{123},ms_{123}}C_{s_{12},0;(k_3-t_3),0}^{s_{123},0} Y_{t_{123},mt_{123}}(\boldsymbol{\lambda}_{c1}) Y_{s_{123},ms_{123}}(\boldsymbol{\rho}_{c2}) Y_{q_{123},mq_{123}}(\textbf{R}_{c3}),
 \end{align}
\end{widetext}
where it becomes evident that the original 9 spherical harmonics (Eq.~\ref{Rab}) are reduced to 3 after undergoing coupling. The detailed procedure involves initially coupling $Y_{t_1,m_{t_1}}(\boldsymbol{\lambda}_{c_1})$ and $Y_{t_2,m_{t_2}}(\boldsymbol{\lambda}_{c_1})$ into $Y_{t_{12},m_{t_{12}}}(\boldsymbol{\lambda}_{c_1})$, accompanied by the relevant CG coefficients $C_{t_1,mt_1;t_2,mt_2}^{t_{12},mt_{12}}$, $C_{t_1,0;t_2,0}^{t_{12},0}$, and normalization factor $\left[ \frac{(2t_1+1)(2t_2+1)}{4\pi(2t_{12}+1)} \right]^{1/2}$. Here, $t_{12}$ and $m_{t_{12}}$ are novel variables derived from coupling $t_1$ and $t_2$. Subsequently, the newly obtained $Y_{t_{12},m_{t_{12}}}(\boldsymbol{\lambda}_{c_1})$ couples with the original $Y_{t_3,m_{t_3}}(\boldsymbol{\lambda}_{c_1})$ to yield $Y_{t_{123},m_{t_{123}}}(\boldsymbol{\lambda}_{c_1})$, accompanied by the relevant CG coefficients $C_{t_{12},mt_{12};t_3,mt_3}^{t_{123},mt_{123}}$, $C_{t_{12},0;t_3,0}^{t_{123},0}$, and normalization factor $\left[ \frac{(2t_{12}+1)(2t_3+1)}{4\pi(2t_{123}+1)} \right]^{1/2}$, where $t_{123}$ and $m_{t_{123}}$  derive from coupling $t_{12}$ and $t_3$. Likewise, we couple the three spherical harmonics $Y_{k_1-t_1,mk_1- mt_1}(\boldsymbol{\rho}_{c_2})$, $Y_{k_2-t_2,mk_2- mt_2}(\boldsymbol{\rho}_{c_2})$, and $Y_{k_3-t_3,mk_3- mt_3}(\boldsymbol{\rho}_{c_2})$ associated with the same motion $\rho_{c_2}$ but with different quantum numbers, resulting in a new spherical harmonic $Y_{s_{123},ms_{123}}(\boldsymbol{\rho}_{c_2})$. The process involves introducing new variables $s_{12},ms_{12}$, $s_{123},ms_{123}$ obtained from the coupling of variables $k_1-t_1,mk_1- mt_1$, $k_2-t_2,mk_2- mt_2$, and $k_3-t_3,mk_3- mt_3$. Similarly, the three spherical harmonics $Y_{L_{a1}-k_1,mL_{a1}- mk_1}(\textbf{R}_{c_3})$, $Y_{L_{a2}-k_2,mL_{a2}- mk_2}(\textbf{R}_{c_3})$, and $Y_{L_{a3}-k_3,mL_{a3}- mk_3}(\textbf{R}_{c_3})$ associated with the same motion $R_{c_3}$ but different quantum numbers are coupled to produce the new spherical harmonic $Y_{q_{123},mq_{123}}(\boldsymbol{R}_{c_3})$. Here, ${q_{12},mq_{12}}$ and ${q_{123},mq_{123}}$ represent the orbital quantum numbers resulting from the coupling of these three spherical harmonics.

\section{Results and discussions}

In this part, we will discuss the equivalence of the two cases:\textcircled{1} The color octet state can be replaced by a superposition of color singlet ground and excited states. \textcircled{2} In the case of a complete basis, diquarks and dimesons are equivalent.

 \begin{table}[th]
\centering
\caption{Results of the coupling of two ground-state channels in $T_{cc}$ system. (unit: MeV)\label{Tcc}}
\begin{tabular}{cccccccccc}
\hline \hline
$|[LS]_i F_j C_k\rangle$&Channel&$E$&$E^{Theo}_{th}$&$E_{B}$ & $E^{Exp}_{th}$\\
\hline
$|5,1,1\rangle$ & $DD^{*}$-$S$~wave            & 3874.12 & 3872.71 &0&  3871.69          \\
$|8,1,1\rangle$ & $D^{*}D^{*}$-$S$~wave        & 3961.50 & 3960.74 &0&  4013.70          \\
\multicolumn{2}{c}{$E_0=E(DD^{*}+D^{*}D^{*}$) }    & 3871.59 & 3872.71 &1.12                 \\
\hline
\end{tabular}
\end{table}

\begin{table}[th]
\centering
\caption{Comparison between the results of the accumulation approach and the complete-channel coupling calculations. (unit: MeV)\label{Tcc2}}
\begin{tabular}{cccccccccc}
\hline \hline
Index&Channel&$E$\\
\hline
5 & $DD^{*}$-$S$~wave                     & 3874.12  \\
8 & $D^{*}D^{*}$-$S$~wave                 & 3961.50           \\
1 & $D^{*}D_1^{\prime}$-$P$~wave          & 4432.14         \\
2 & $D^{*}D_1^{*}(P)$-$P$~wave            & 4419.64         \\
3 & $D_{2}^{\prime}D_{2}^{*}(P)$-$P$~wave & 5360.48         \\
4 & $D_{2}^{*}(D)D_{2}^{*}(P)$-$P$~wave   & 5379.45         \\
\multicolumn{2}{c}{Complete-coupled-channels:}                   & 3871.19                  \\
\hline
Index&Channel&$\Delta E_i$ \\
\hline
1 & $DD^{*}+D^{*}D^{*}+D^{*}D_1^{\prime}$-$P$~wave             &-0.21        \\
2 & $DD^{*}+D^{*}D^{*}+D^{*}D_1^{*}(P)$-$P$~wave               &-0.12        \\
3 & $DD^{*}+D^{*}D^{*}+D_{2}^{\prime}D_{2}^{*}(P)$-$P$~wave    &-0.04        \\
4 & $DD^{*}+D^{*}D^{*}+D_{2}^{*}(D)D_{2}^{*}(P)$-$P$~wave      &-0.03        \\
\multicolumn{2}{c}{$E_0$+$\sum\limits_{i=1}^4 \Delta E_i$:}&3871.17\\
\hline
\end{tabular}
\end{table}
 In the chiral quark model, $T_{cc}$ system is special, because $DD^{*}$ and $D^{*}D^{*}$ are coupled to form a bound state, which is denoted as $E_0$, shown in Table \ref{Tcc}. However, the expansion of the wave function of the highly excited state is too complicated for us (see Eq.~\ref{Y_Ex}), and the number of channels of the wave function of the highly excited state is very large. In principle, the bound states of a system should be calculated directly by calculating the complete-channel coupling of all channels. In this article, we take an accumulation approach.  Since the two channels $DD^{*}$ and $D^{*}D^{*}$ can already form a bound state, we coupled all highly excited physical channels with $DD^{*} + D^{*}D^{*}$ ($E_0$) respectively to find the contribution of each channel to the bounding energy of the system, which can be expressed by $\Delta {E_i}= E(DD^{*} + D^{*}D^{*} + ith~channel) - E_0$. Finally, The total energy of the system can be expressed as the coupling energy of $E_0$ plus all high excited state's contribution ($\Delta {E_i}$) to the system of binding energy. Its calculation formula of the accumulation approach  can be expressed as:
\begin{align}
\label{Ei}
 E = E_0 + \sum\limits_{i=1}^n \Delta {E_i}.
\end{align}
To assess the accuracy of the accumulation approach, we performed calculations for the six channels in the $T_{cc}$ system, which involved both direct calculation with complete channel couplings and calculation using the accumulation approach (see in Table \ref{Tcc2}). The six channels include two states with $S$-wave ($DD^{*}$ and $D^{*}D^{*}$) and four states with $P$-wave ($D^{*}D_1^{\prime}$, $D^{*}D_1^{*}(P)$, $D_{2}^{\prime}D_{2}^{*}(P)$, and $D_{2}^{*}(D)D_{2}^{*}(P)$). The results of the complete-channel coupling calculations indicate that the lowest eigenenergy for these six channels in the $T_{cc}$ system is $3871.19$ MeV. According to the accumulation approach, the coupling energy for the channels $DD^{*}$ and $D^{*}D^{*}$ is $3871.59$ MeV ($E_0$=$3871.59$ MeV in Table \ref{Tcc}). The coupling of the first channel $D^{*}D_1^{\prime}$ with ($DD^{*}$ and $D^{*}D^{*}$) reduces the energy of $E_0$ by 0.21 MeV, denoted as $\Delta E_1$. Similarly, the coupling of the second channel $D^{*}D_1^{*}(P)$ further lowers the energy of $E_0$ by 0.12 MeV, denoted as $\Delta E_2$. The contributions to the coupling energy from $D^{*}D_1^{*}(P)$ and $D_{2}^{\prime}D_{2}^{}(P)$ are denoted as $\Delta E_3$ and $\Delta E_4$, respectively. Finally, utilizing the calculation formula in the accumulation approach, $E = E_0 + \sum\limits_{i=1}^4 \Delta {E_i}$, the calculated final energy is $3871.17$ MeV. This implies that the accuracy of the accumulation approach is applicable in the calculation of bound states. Therefore, in this paper, we employ the accumulation approach for the calculation of the remaining channels.

\subsection{Equivalence between color octet and orbital excited state}
For the $T_{cc}$ system, there are two ground-state molecular states: $DD^{*}$and $D^{*}D^{*}$. They have two corresponding color octet, $[D]_8[D^*]_8$ and $[D^*]_8[D^*]_8$. Here, we consider the equivalence between higher-order partial wave and color octet, so the color octet here only considers the two ground states, and the excited color octet is not taken into account. As shown in Table \ref{try}, the contributions of $[D]_8[D^{*}]_8$ and $[D^{*}]_8[D^{*}]_8$ to the binding energy of the system are $\Delta E_{28}=-0.37$ MeV and $\Delta E_{31}=-0.56$ MeV, respectively. According to the calculation Eq.~\ref{Ei} of the accumulation approach, the total energy is $E_1=E_0+\Delta E_{28}+\Delta E_{29}=3870.66$ MeV.
\begin{table*}[th]
  \centering
  \fontsize{10}{9}\selectfont
  \makebox[\textwidth][c]{
   \begin{threeparttable}
   \caption{Contributions to the binding energy from all possible channels in the $T_{cc}$ system.}\label{try}
    \begin{tabular}{ccccccccc}
    \hline\hline
Index(i)&Channel&$\Delta E_i$ & Index(i)&Channel&$\Delta E_i$ \\\hline
1  & $DD^{*}+D^{*}D^{*}+D^{*}D_1^{\prime}$-$P$~wave              &-0.21 &24 & $DD^{*}+D^{*}D^{*}+[D^{*}D_{1}^{'}]_8$-$P$~wave           & -0.17        \\
2  & $DD^{*}+D^{*}D^{*}+D^{*}D_1^{*}(P)$-$P$~wave                &-0.12 &25 & $DD^{*}+D^{*}D^{*}+[D^{*}D_{1}^{*}(P)]_8$-$P$~wave        & -0.34        \\
3  & $DD^{*}+D^{*}D^{*}+D_{2}^{\prime}D_{2}^{*}(P)$-$P$~wave     &-0.04 &26 & $DD^{*}+D^{*}D^{*}+[D_2^{'}D_{2}^{*}(P)]_8$-$P$~wave      & -0.02        \\
4  & $DD^{*}+D^{*}D^{*}+D_{2}^{*}(D)D_{2}^{*}(P)$-$P$~wave       &-0.03 &27 & $DD^{*}+D^{*}D^{*}+[D_2^{*}(D)D_2^{*}(P)]_8$-$P$~wave     & -0.10        \\
5  & $DD^{*}+D^{*}D^{*}+D_{0}^{*}D_{1}^{\prime}(P)$-$S$~wave     &-0.31 &28 & $DD^{*}+D^{*}D^{*}+[DD^{*}]_8$-$S$~wave                   & -0.37        \\
6  & $DD^{*}+D^{*}D^{*}+D_{0}^{*}D_{1}^{*}(P)$-$S$~wave          &-1.03 &29 & $DD^{*}+D^{*}D^{*}+[D_0^{*}D_1^{'}(P)]_8$-$S$~wave        & -0.57        \\
7  & $DD^{*}+D^{*}D^{*}+D_{}^{*}D_{1}^{*}(D)$-$S$~wave           &-0.01 &30 & $DD^{*}+D^{*}D^{*}+[D_0^{*}D_1^{*}(P)]_8$-$S$~wave        & -0.67        \\
8  & $DD^{*}+D^{*}D^{*}+D_{1}^{\prime}D_{1}^{\prime}$-$S$~wave   &-0.01 &31 & $DD^{*}+D^{*}D^{*}+[D^{*}D^{*}]_8$-$S$~wave               & -0.56        \\
9  & $DD^{*}+D^{*}D^{*}+D_{1}^{\prime}D_{1}^{*}(P)$-$S$~wave     &-0.25 &32 & $DD^{*}+D^{*}D^{*}+[D^{*}D_1^{*}(D)]_8$-$S$~wave          & -0.01        \\
10 & $DD^{*}+D^{*}D^{*}+D_{1}^{*}(P)D_{1}^{*}(P)$-$S$~wave       &-0.11 &33 & $DD^{*}+D^{*}D^{*}+[D_1^{'}D_1^{'}]_8$-$S$~wave           & -0.00        \\
11 & $DD^{*}+D^{*}D^{*}+D_{1}^{*}(D)D_{1}^{*}(D)$-$S$~wave       &-0.01 &34 & $DD^{*}+D^{*}D^{*}+[D_1^{'}D_1^{*}(P)]_8$-$S$~wave        & -1.05        \\
12 & $DD^{*}+D^{*}D^{*}+D_{0}^{*}D_{1}^{\prime}$-$D$~wave        &-0.09 &35 & $DD^{*}+D^{*}D^{*}+[D_1^{*}(P)D_1^{*}(P)]_8$-$S$~wave     & -0.45        \\
13 & $DD^{*}+D^{*}D^{*}+D_{0}^{*}D_{1}^{*}(P)$-$D$~wave          &-0.05 &36 & $DD^{*}+D^{*}D^{*}+[D_1^{*}(D)D_1^{*}(D)]_8$-$S$~wave     & -0.01        \\
14 & $DD^{*}+D^{*}D^{*}+D_{}^{}D_{}^{*}$-$D$~wave                &-0.54 &37 & $DD^{*}+D^{*}D^{*}+[D_0^{*}D_{1}^{'}]_8$-$D$~wave         & -0.04        \\
15 & $DD^{*}+D^{*}D^{*}+D_{1}^{\prime}D_{1}^{\prime}$-$D$~wave   &-0.01 &38 & $DD^{*}+D^{*}D^{*}+[D_0^{*}D_{1}^{*}(P)]_8$-$D$~wave      & -0.02        \\
16 & $DD^{*}+D^{*}D^{*}+D_{1}^{\prime}D_{1}^{*}(P)$-$D$~wave     &-0.01 &39 & $DD^{*}+D^{*}D^{*}+[DD^{*}]_8$-$D$~wave                   & -0.02        \\
17 & $DD^{*}+D^{*}D^{*}+D_{}^{*}D_{}^{*}$-$D$~wave               &-0.02 &40 & $DD^{*}+D^{*}D^{*}+[D_1^{'}D_{1}^{'}]_8$-$D$~wave         & -0.00        \\
18 & $DD^{*}+D^{*}D^{*}+D_{2}^{\prime}D_{}^{*}$-$D$~wave         &-0.29 &41 & $DD^{*}+D^{*}D^{*}+[D_1^{'}D_{1}^{*}(P)]_8$-$D$~wave      & -0.06        \\
19 & $DD^{*}+D^{*}D^{*}+D_{2}^{*}D_{}^{*}$-$D$~wave              &-0.04 &42 & $DD^{*}+D^{*}D^{*}+[D_1^{*}(P)D_{1}^{*}(P)]_8$-$D$~wave   & -0.05        \\
20 & $DD^{*}+D^{*}D^{*}+D_{2}^{*}(D)D_{}^{*}$-$D$~wave           &-0.05 &43 & $DD^{*}+D^{*}D^{*}+[D^{*}D^{*}]_8$-$D$~wave               & -0.01        \\
21 & $DD^{*}+D^{*}D^{*}+D_{2}^{*}(P)D_{1}^{\prime}$-$D$~wave     &-0.02 &44 & $DD^{*}+D^{*}D^{*}+[D_2^{'} D^{*}]_8$-$D$~wave            & -0.06        \\
22 & $DD^{*}+D^{*}D^{*}+D_{2}^{*}(P)D_{1}^{*}(P)$-$D$~wave       &-0.01 &45 & $DD^{*}+D^{*}D^{*}+[D_2^{*}(D)D^{*}]_8$-$D$~wave          & -0.17        \\
23 & $DD^{*}+D^{*}D^{*}+D_{2}^{*}(P)D_{2}^{*}(P)$-$D$~wave       &-0.03 &46 & $DD^{*}+D^{*}D^{*}+[D_2^{*}(P)D_1^{'}]_8$-$D$~wave        & -0.00        \\
                                                                      &&&47 & $DD^{*}+D^{*}D^{*}+[D_2^{*}(P)D_{1}^{*}(P)]_8$-$D$~wave   & -0.03        \\
                                                                      &&&48 & $DD^{*}+D^{*}D^{*}+[D_2^{*}(P)D_{2}^{*}(P)]_8$-$D$~wave   & -0.01        \\
\multicolumn{6}{c}{$E_1$=$E_0$+$\Delta E_{28}$+$\Delta E_{29}$=3870.66}     \\
\multicolumn{6}{c}{$E_2$=$E_0$+$\sum\limits_{i=1}^{23} \Delta E_i$=3868.31} \\
\multicolumn{6}{c}{$E_3$=$E_0$+$\sum\limits_{i=1}^{48} \Delta E_i$=3863.75:}\\
\hline
    \end{tabular}
   \end{threeparttable}}
  \end{table*}
For the higher order component of $T_{cc}$ system, we only consider three cases: the quantum numbers of $J_{12}$ of the two clusters are 0, 1, and 2, and the orbital quantum numbers of the intercluster motion $L_3$ are 1, 0, and 2, respectively. Other $J_{12} \times L_{3}$ combinations are not considered in this article due to computational complexity. For the first case $J_{12}=0$, $L_{3}=1$, we can see that in the table, there are only 4 physical channels with intergroup orbits are $P$-wave due to the limitation of symmetry. They are $D^{*} D_1^{\prime}$, $D^{*} D_1^{*}$, $D_2^{\prime} D_2^{*}$ and $D_2^{*}(D) D_2^{*}(P)$. The two physical channels, $D^{*} D_1^{\prime} $ and $D^{*} D_1^{*} $, contribute relatively large to the bound energy of the $T_{cc} $ system, about -0.1 MeV. While the other two physical channels, $D_2^{\prime} D_2^{*} $ and $D_2^{*}(D) D_2^{*}(P)$, because they are highly excited states, their contribution to the bound energy of the $T_{cc}$ system is much smaller, about -0.03 MeV.  For the second case of $J_{12}=1$, $L_{3}=0$, there are only seven physical channels with $S$-wave between the clusters due to the symmetry. Also due to the $S$-wave, their contribution to the binding energy of the $T_{cc}$ system is obviously larger. The largest contribution comes from the $D_0^{*} D_1^{*}(P)$, which reaches -1.03 MeV. The contribution of the two physical channels containing $D$-wave, $D^{*} D_1^{*}(D)$ and $D_1^{*}(D) D_1^{*}(D)$, is almost zero. The contribution of other channels is within -0.3 MeV. In the third case, $J_{12} =2$, $L_{3} =2$, there are 12 different physical channels with $D$-wave between the clusters. Because they are $D$-wave between the clusters, most of the physical channels have relatively small contributions to the bound state of the $T_{cc}$ system, with absolute values less than 0.1 MeV. But there are two physical channels with large contributions, which are $D$-wave excitation of $DD^{*}$ and $D^{*}D^{*}$ . Because of the large $S$-$D$ coupling effect, they respectively contribute -0.54 MeV and -0.29 MeV to the bound state of the $T_{cc}$ system. The 12 physical channels with $D$-wave between the clusters contribute a total of -1.15 MeV to the bound state of the $T_{cc} $ system. Combined with the contributions of all the 7 physical channels with S wave between the clusters, 4 physical channels with P wave between the clusters, and 12 physical channels with $D$-wave between the clusters, the lowest energy of the $T_{cc}$ system can be suppressed to 3868.31 MeV ($E_2$=$E_0$+$\sum_{i=1}^{23} \Delta E_i$=3868.31 MeV).

For the contribution of the color-octet state, we only consider two channels, namely, $[DD^{*}]_8$ and $[D^{*} D^{*}]_8$. The contribution of the two color-octet channels is relatively large, reaching -0.37 MeV and -0.56 MeV respectively. Combined with the original color-singlet energy of $DD^{*} +D^{*}D^{*}$, the energy of the system is $T_{cc} $ is 3870.66 MeV. We can find that the contribution of the high-order excited states  and the color-octet states are almost the same. It is worth noting that the contribution of the two color-octet channels, $[DD^{*}]_8$ and $[D^{*} D^{*}]_8$, and the contribution of  $D$-wave of  $DD^{*}$ and $D^{*}D^{*} $ is almost the same, which is about -0.9 MeV. Therefore, we believe that the contribution of the color-octet channel can be replaced by the high-order excited states in the molecular state.

\subsection{Equivalence between diquark structure and molecular structure}

The $T_{cc}$ system has two significant diquark channels, namely $\bar{3}\times 3$ and $6\times \bar{6}$. In the quark models, the energy associated with $\bar{3}\times 3$ typically lies below the threshold $DD^{*}$. In this study, it is observed that $\bar{3}\times 3$ possesses an intrinsic energy of 3709 MeV, representing a bound state with a binding energy of 163 MeV in Table \ref{Tcc3}. After channel coupling, the coupling energy decreases to 3705 MeV and its total binding energy increases to 167 MeV.
 \begin{table}[th]
\centering
\caption{Calculation results for the diquark structure. (unit: MeV)\label{Tcc3}}
\begin{tabular}{cccccccccc}
\hline \hline
$|[LS]_i F_j C_k\rangle$&Channel&$E$&$E^{Theo}_{th}$&$E_{B}$ & $E^{Exp}_{th}$\\
\hline
$|5,2,4\rangle$ & $[cc]_6^0[\bar{q}\bar{q}]_{\bar{6}}^1$-$S$~wave            & 4134.6 & 3872.71 &0&  3871.69          \\
$|8,2,3\rangle$ & $[cc]_{\bar{3}}^1[\bar{q}\bar{q}]_{3}^1$-$S$~wave          & 3709.3 & 3872.71 &163.4&  3871.69          \\
\multicolumn{2}{c}{$E([cc]_6^0[\bar{q}\bar{q}]_{\bar{6}}^1+[cc]_{\bar{3}}^1[\bar{q}\bar{q}]_{3}^1$) }    & 3705.4 & 3872.71 &167.3                 \\
\hline
\end{tabular}
\end{table}

In the molecular structure of $T_{cc}$, there are 25 color singlet states and 25 corresponding color octet states. They can be categorized based on the difference in inter-subgroups orbital functions: $S$-wave, $P$-wave, and $D$-wave. As discussed previously, the channel coupling result for all color singlet states is 3868.31 MeV with a binding energy of 3.8 MeV.
Therefore,  we only need to consider the contributions of all $S$-, $P$-, and $D$-wave color octet states. The contribution from four color octet states with $P$ wave orbital function is $\Delta E_{24}=-0.17$ MeV, $\Delta E_{25}=-0.34$ MeV, $\Delta E_{26}=-0.02$ MeV, $\Delta E_{27}=-0.10$ MeV respectively.Indices from 27 to 36 in Table \ref{try} correspond to the contributions of $S$-wave color octet states. Due to the $S$-wave nature of inter-cluster motion, the contributions of these nine channels are relatively bigger. With the exceptions of $[D^{*} D_1^{*}(D)]_8$, $[D_1^{'} D_1^{'}]_8$, and $[D_1^{*}(D) D_1^{*}(D)]_8$, whose contributions to the binding energy are approximately zero, the remaining six channels contribute to the system's binding energy in the range of -0.4 to -1.0 MeV. The contribution from color octets with $D$ wave orbital function has little impact on the binding energy, as their absolute value for $\Delta E_i$ is less than 0.2 MeV. Finally, the lowest energy of the $T_{cc}$ system is finally determined to be $E_{3}$=3863.75 MeV, taking into account all excited color singlets and color octet molecular channels.

According to the aforementioned discussion, it is challenging to conclude that the diquark structure is equivalent to the molecular state structure. There are two potential reasons for this assertion. Firstly, due to the intricacy of calculations, it becomes arduous to consider additional molecular state excitation channels in these computations. Although highly excited states' physics contributes minimally to bound states of the system, a significant contribution always exists when there is a large number of highly excited states. However, we do not believe that relying on highly excited states' physics can sufficiently reduce the minimum energy of the $T_{cc}$ system close to that of diquarks. Secondly, while equivalence between diquark and molecular state structures is derived from a wave function perspective, determining direct parameters for diquarks in Hamiltonian remains unfeasible. This discrepancy arises because parameters in Hamiltonian for molecular states can be directly determined based on experimental discoveries related to mesons; however, such determination cannot be applied directly towards establishing parameters for diquarks in Hamiltonian as they only pertain to molecular state parameters. All these factors may lead us towards concluding that both diquark and molecular structures are indeed equivalent.

\section{Summary}

 In the framework of the chiral quark model, we have investigated the equivalence between different quark structures in our system with help of Gaussian expansion method. This includes exploring whether the influence of color octet states on the ground state can be replaced by high-excited states of color singlet, and the equivalence between diquark and dimeson structures. Due to the complexity in handling the wave functions of highly excited states, the accumulation approach is employed in this paper to calculate the binding energies of various structures.

 The calculation results show that the energies coupled for the two $S$-wave states, $DD^{*}$ and $D^{*} D^{*}$, and their corresponding color-octet states, $[DD^{*}]_8$ and $[D^{*} D^{*}]_8$, amount to 3870.66 MeV. When considering all possible color-singlet excited states, the energy is calculated to be 3868.31 MeV. The results from these two cases are approximately equal, implying that the impact of color-octet states on the ground state is comparable to that of color-singlet excited states.

 The lowest energy of the diquark structure in the $T_{cc}$ system is 3705.4 MeV, with a binding energy of 167 MeV. We considered all possible excited states in the dimeson structure, including color singlet and color octet states. The calculational results show a binding energy of 3863.7 MeV. Although the binding energy is further reduced, it is still significantly higher than the energy of the diquark structure. This result could stem from the limited consideration of highly excited states. Additionally, it is possible that errors may arise in the calculation of the diquark structure because all parameters in the Hamiltonian are directly determined by the meson spectrum.

 In summary, we believe that in the dimeson structure, the contribution of color octet states can be replaced by highly excited states of color singlet. Despite considering a sufficient number of bases, the diquark structure cannot be entirely replaced by the dimeson structure.

\acknowledgments{ This work is supported partly by the National Natural Science Foundation of China under Grant Nos. 12205249, 12205125, 11675080, 11775118 and 11535005 by the Natural Science Foundation of Jiangsu Province (BK20221166), and the Funding for School-Level Research Projects of Yancheng Institute of Technology (No. xjr2022039).}

\end{document}